\documentclass[11pt,a4paper]{article}

\usepackage{amsmath,amsthm,amssymb}
\usepackage{mathtools}

\usepackage{fullpage}

\usepackage{color}

\usepackage{graphicx} 

\usepackage[colorlinks=true,citecolor=blue,linkcolor=red,urlcolor=blue]{hyperref}
\usepackage[font=footnotesize, labelfont=bf, margin=0.5cm]{caption}
\usepackage[labelformat=simple]{subcaption}

\newtheorem{lem}{Lemma}

\newcommand{\tube}{\mathbb T}
\newcommand{\tubeLM}{\mathbb T_{L,M}}

\newcommand{\betafull}{\beta^\mathrm{F}}
\newcommand{\pham}{p^\mathrm{H}}
\newcommand{\kappaham}{\kappa^\mathrm{H}}
\newcommand{\muham}{\mu^\mathrm{H}}
\newcommand{\chiham}{\chi^\mathrm{H}}
\newcommand{\nuham}{\nu^\mathrm{H}}

\newcommand{\M}{\mathbf{M}}

\title{Knotting statistics for polygons in lattice tubes}

\author{N.~R.~Beaton\\
\small School of Mathematics and Statistics\\
\small The University of Melbourne, Victoria 3010, Australia
\and
J.~W.~Eng \hspace{1cm} C.~E.~Soteros\thanks{\href{mailto:soteros@math.usask.ca}{soteros@math.usask.ca}}\\
\small Department of Mathematics and Statistics\\
\small The University of Saskatchewan, Saskatoon, SK, Canada S7N 5E6
}

\RequirePackage[maxbibnames=99,backend=bibtex,sorting=none]{biblatex}
\renewbibmacro{in:}{. In:}
\DeclareFieldFormat[misc]{title}{\mkbibquote{#1}}

\DeclareFieldFormat[article]{volume}{\mkbibbold{#1}}
\DeclareFieldFormat{url}{\textsc{url}: \href{#1}{#1}}
\DeclareFieldFormat{doi}{\textsc{doi}: \href{http://dx.doi.org/#1}{#1}}
\DeclareFieldFormat{eprint:arxiv}{arXiv:\href{https://arxiv.org/abs/#1}{#1}}

\begin{document}

\maketitle

\begin{abstract}
We study several related models of self-avoiding polygons in a tubular subgraph of the simple cubic lattice, with a particular interest in the asymptotics of the knotting statistics. Polygons in a tube can be characterised by a finite transfer matrix, and this allows for the derivation of pattern theorems, calculation of growth rates and exact enumeration. We also develop a static Monte Carlo method which allows us to sample polygons of a given size directly from a chosen Boltzmann distribution.

Using these methods we accurately estimate the growth rates of unknotted polygons in the $2\times1\times\infty$ and $3\times1\times\infty$ tubes, and confirm that these are the same for any fixed knot-type $K$. We also confirm that the entropic exponent for unknots is the same as that of all polygons, and that the exponent for fixed knot-type $K$ depends only on the number of prime factors in the knot decomposition of $K$. For the simplest knot-types, this leads to a good approximation for the polygon size at which the probability of the given knot-type is maximized, and in some cases we are able to sample sufficiently long polygons to observe this numerically.

\end{abstract}

\begin{center} Dedicated to Stuart Whittington on the occasion of his 75\textsuperscript{th} birthday.\end{center}

\section{Introduction}\label{sec:intro}

Self-avoiding walk and self-avoiding polygon models are the standard statistical mechanics lattice models for linear and ring polymers in dilute solutions \cite{vanRensburg2015Statistical,vanderzande_1998}.
They are also of interest due to their connection with critical phenomena for other lattice models of statistical mechanics \cite{ISI:000267081500016,StuReview82}.  Initial studies of self-avoiding polygon models focused on studying polygon counts and average geometric properties (e.g.~radius of gyration) via exact enumeration, theoretical analysis and Monte Carlo sampling.  For example, early contributions of S.~G.~Whittington to the field contributed in all three of these aspects \cite{GuttmannWhittington_1978,WhittingtonValleauJCP69}.  Subsequently, interest in studying the average topological properties of polygons has grown.  This interest arose from the seminal paper by Sumners and Whittington \cite{SumnersWhittington_1988}, where they used a lattice polygon model to establish the long-standing Frisch-Wasserman-Delbruck (FWD) \cite{Delbruck62,Frisch61} conjecture that sufficiently long ring polymers have a high probability of being knotted.  Continued interest related to this has been motivated by a concomitant growth in experimental data related to DNA topology (see for example the review articles~\cite{lim_molecular_2015,orlandini_statics_2018} and references therein).  

Early numerical studies of knotting probabilities for lattice polygons were due to Janse van Rensburg and Whittington \cite{vanRensburgWhittington_90} using Markov chain Monte Carlo simulations of fixed-length polygons on the fcc lattice.  Some corresponding off-lattice results were subsequently obtained by Katrich et al.~\cite{Katrich2000} as well as Shimamura and Deguchi \cite{SHIMAMURA2000184}.  From such off-lattice studies \cite{SHIMAMURA2000184,0953-8984-27-35-354104,doi:10.1063/1.4996645}, Deguchi and co-workers proposed a general asymptotic form for the probability of a given knot-type $K$  as a function of polygon length $n$:
\begin{equation}
\mathbb{P}_n(K)\sim A_K (n-\Delta N(K))^{m(K)} \exp\left(-\frac{n-\Delta N(K)}{N_K}\right), \qquad  n \rightarrow \infty,
\label{deguchi-form}
\end{equation}
where $A_K, \Delta N(K), m(K)$ and $N_K$ are expected to be constants which may depend on $K$ but not $n$. 
This form is consistent with lattice evidence \cite{PhysRevE.86.031805,1742-5468-2014-2-P02014,JansevanRensburg2008,1751-8121-44-16-162002,0305-4470-31-28-010,0305-4470-34-37-310} that indicates that $p_n(K)$, the number of $n$-edge self-avoiding polygons (counted up to translation) with fixed knot-type $K$,  has the following asymptotic form:
\begin{equation}
p_n(K)\sim C_K \mu _{0_1} ^n n^{\alpha_{0_1} -3+f_K}, \qquad {\rm as}\quad n \rightarrow \infty, 
\label{pnKasymptotics}
\end{equation}
where $f_{0_1}=0$ and otherwise $f_K$ is the number of prime knot factors ({\it factors}, for short) in the prime knot decomposition of knot-type $K$ and where $\mu_{0_1}$ is the lattice-dependent {\it growth constant} for unknotted polygons ($\mu_{0_1}$ is proved to exist in \cite{SumnersWhittington_1988}). For all polygons, the conjectured asymptotics are
\begin{equation}
p_n=\sum_{K} p_n(K) \sim C \mu^n n^{\alpha -3}, \qquad {\rm as}\quad n \rightarrow \infty, 
\label{pnasymptotics}
\end{equation}
with the growth constant $\mu >  \mu _{0_1}$ (this strict inequality is established in the proof of the FWD conjecture \cite{SumnersWhittington_1988}). For lattice polygons, there is evidence that the unknot entropic exponent $\alpha_{0_1}=\alpha$; hence in the case that each $n$-edge lattice polygon is considered to be equally likely,  (\ref{pnKasymptotics}) and (\ref{pnasymptotics}) lead to a conjectured asymptotic form consistent with (\ref{deguchi-form}):
\begin{align}
\mathbb{P}_n(K) &= \frac{p_n(K)}{p_n} \notag\\
&\sim A_K (n-\Delta N(K))^{f_K} \exp[-(n-\Delta N(K))(\log\mu -\log \mu_{0_1})],
\label{probKasymptotics}
\end{align}
with $m(K)=f_K$ and $N_K=N_{0_1}=(\log\mu -\log \mu_{0_1})^{-1}$.
There is also numerical evidence from lattice models that the 
amplitude ratios $C_{K_1}/C_{K_2}=A_{K_1}/A_{K_2}$ for different prime knots $K_1$ and $K_2$ are lattice-independent  (i.e.~universal) \cite{1751-8121-44-16-162002}. However,  recent evidence for an off-lattice model (which allows for varying a cylindrical radius of excluded volume) indicates that these amplititude ratios may depend on the extent (the radius) of excluded-volume in the model \cite{doi:10.1063/1.4996645}. 
Without excluded volume,  numerical results~\cite{stutalk2018} for equilateral off-lattice polygons support the following form for $\mathbb{P}_n(0_1)$,
\begin{equation}
\mathbb{P}_n(0_1)\sim A_{0_1} n^{m(0_1)} e^{-n/N_{0_1}}\left(1+\frac{B}{n^{\Delta}}+\frac{C}{n}\right), \qquad  n \rightarrow \infty,
\label{stu-form}
\end{equation}
with estimates for  $\Delta\approx 1/2$ and $m(0_1)\approx -0.125$, indicating that $m(0_1)<0$ for this model while for lattice polygons $m(0_1)$ is expected to be 0. 


Using the form of equation (\ref{probKasymptotics}), for any fixed $K$, $\mathbb{P}_n(K)$ decays exponentially to zero as $n\to\infty$ but for any $K\neq 0_1$,  
$\mathbb{P}_n(K)$ will increase initially with $n$, reaching a maximum at $n\approx \Delta N(K) +f_KN_{0_1}$ before decaying.  
Consistent with this, numerical evidence from various lattice and off-lattice models indicates that for a given model and for the simplest knots, $N_{0_1}\gg\Delta N(K)$ and the knot-type dependence of the  location of the maximum depends primarily on $f_K$, the number of prime knot factors in the prime-knot decomposition of $K$.   Note however that  $\Delta N(K)$ is expected to depend on the minimum number of edges needed to create the knot, and hence it is generally non-zero, dependent on $K$ and does contribute to the location of the maximum.  

Parallel to the interest in polymer entanglement complexity, there has also been interest in the properties of lattice models of polymers in confined geometries (see for example  reviews in \cite{StuEnzo,Micheletti_2011,vanRensburg2015Statistical,orlandini_statics_2018}).  The nature of the confinement considered has ranged from  partial confinement in wedges, slabs or tubes (prisms), where the polymer is allowed to extend freely in one or more direction, to full confinement in a sphere or box.  Initial studies \cite{Whittington1983,HW1985,SW1988,SW1991}  focused on the effects of confinement on limiting free energies, entropic exponents and geometric properties.  Recent interest has turned to studying the effects on knotting statistics.
Respective versions of the FWD conjecture have been proved for wedges \cite{SSW1999}, slabs \cite{Tesi-slab-1994,Tesi1998} and tubes \cite{Soteros1998}, and knotting statistics are believed to follow asymptotic forms similar to those given above.  Other studies have considered the effect of confinement on the ``size" of the knotted part and the extent of localization of the knot \cite{BeatonSoftMatter_2018}.     

In general, for most lattice polygon models, little can be proved about the asymptotic forms (\ref{pnKasymptotics}), (\ref{pnasymptotics}), (\ref{probKasymptotics}) beyond the existence of growth constants for all and unknotted polygons and the corresponding FWD conjecture proof. However, in the case of polygons confined to any $L\times M \times \infty$ simple-cubic lattice tube,   it is known \cite{Soteros1998} via transfer matrix arguments that
\begin{equation}
p_{\tube,n} \sim C_{\tube} \mu_{\tube}^n, \qquad {\rm as}\quad n \rightarrow \infty, 
\end{equation}
where $p_{\tube,n}$ counts the number of $n$-edge polygons in a given tube $\tube$ (up to translation in the direction of the infinite axis of the tube) and the constants $C_{\tube}$ and $\mu_{\tube}$ are determined by the eigenvectors and eigenvalues of the transfer matrix.   Very recently, for the smallest such tube that admits non-trivial knots ($2\times 1 \times \infty$),  for knots $K$ in the set of all 2-bridge knots with unknotting number one or knots formed from their connect-sum,  it has been established~\cite{manyauthors_2x1tubes} for $n$ sufficiently large that there exist constants $B_K$ and $D_K$ such that
\begin{equation}
B_K p_{\tube,n}(0_1) n^{f_K} \leq p_{\tube,n}(K)\leq D_K p_{\tube,n}({0_1}) n^{f_K},
\label{pnKtubeinequality1}
\end{equation}
where $p_{\tube,n}(K)$ is the number of polygons counted in $p_{\tube,n}$ that have knot-type $K$.
This establishes the expected form from (\ref{pnKasymptotics}) for the  growth constant and the increase in the entropic exponent.  The arguments used for this, however, do not appear to be easily extended to other knot-types or larger tube sizes.   Consistent with (\ref{pnKasymptotics}) and (\ref{pnKtubeinequality1}), it is conjectured that for any knot $K$, 
 \begin{equation}
 p_{\tube,n}(K) \sim C_{\tube,K} n^{f_K} \mu_{\tube,0_1}^n, \qquad {\rm as}\quad n \rightarrow \infty .
 \label{pnKtubeinequality}
 \end{equation}

In this paper, we explore some of the remaining open questions about the knotting statistics in tubes by using transfer matrix arguments, exact enumeration and Monte Carlo methods for tube sizes $L\times 1 \times \infty$ for $L=2$ and $3$.   In particular, we provide numerical evidence that (\ref{pnKtubeinequality}) holds  and that the entropic exponent for unknots  is the same as that for all polygons, i.e. that there exist constants $C_{\tube,0_1}$ and $\mu_{\tube,0_1}$ consistent with 
 \begin{equation}
 p_{\tube,n}(0_1) \sim C_{\tube,0_1} \mu_{\tube,0_1}^n, \qquad {\rm as}\quad n \rightarrow \infty .
 \label{pnunknottube}
 \end{equation}
Our evidence is primarily based on counts of polygons in tubes enumerated by maximum span $s$ (in the unconfined direction) instead of by number of edges $n$; the asymptotic forms of (\ref{pnKtubeinequality}) and (\ref{pnunknottube}) should not depend on whether $s$ or $n$ is used but the  constants in these equations are expected to change.


The outline of the paper is as follows. In Section~\ref{sec:themodel} we define a general model of self-avoiding polygons confined to a lattice tube, with Boltzmann weights associated with length and/or span along the axis of the tube. In Section~\ref{sec:numerical} we describe how transfer matrices can be used to study this model, and introduce a Monte Carlo algorithm for sampling random polygons directly from the desired Boltzmann distribution.  Section~\ref{sec:results} contains a variety of numerical results, including enumerations and growth rates obtained directly from the transfer matrices, and estimates of knotting probabilities, unknot growth rates, critical exponents and amplitude ratios obtained via simulation. These results are used to provide evidence for the asymptotic forms (\ref{pnKtubeinequality}) and (\ref{pnunknottube}). Finally some concluding remarks are given in Section~\ref{sec:conclusion}.

\section{Theory and exact results}\label{sec:themodel}

To begin we need some definitions.  {For non-negative integers $L,M$, let} $\tubeLM\equiv\tube\subset \mathbb Z^3$ be the semi-infinite $L\times M$ tube on the simple cubic lattice defined by
$$\tube = \{(x,y,z)\in \mathbb Z^3:x\geq0, 0\leq y\leq L, 0 \leq z \leq M\}.$$

\subsection{The fixed-edge model}

We  consider first  a general model for polymers in  tubes where the polymer (modelled by a polygon in a lattice tube) is subject to an external force.  This model has been studied previously -- see \cite{Beaton_2016} and references therein. The notation and definitions used here (unless stated otherwise) are as in \cite{Beaton_2016}.  
Define $\mathcal P_\tube$ to be the set of self-avoiding polygons in $\tube$ which occupy at least one vertex in the plane $x=0$, and let $\mathcal P_{\tube,n}$ be the subset of $\mathcal P_\tube$ comprising polygons with $n$ edges ($n$ even). Then let $p_{\tube,n}  = |\mathcal P_{\tube,n}|$. 

\begin{figure}[h]
\centering
\resizebox{0.6\textwidth}{!}{
\includegraphics[width=\textwidth]{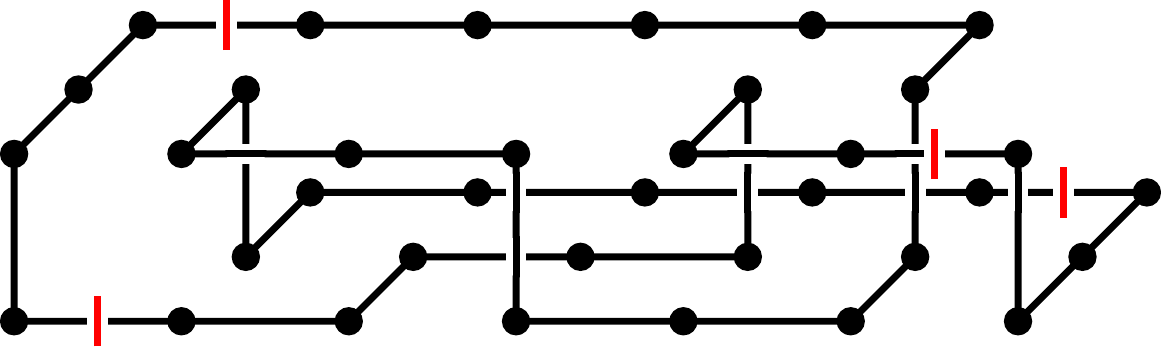}
}
\caption{A 36-edge polygon $\pi$ that fits inside $\tube_{L,M}$ with $L\geq 2$ and $M\geq 1$;  the tube extends without bound to the right and the span $s(\pi)=6$.
The locations of the two pairs of vertical red lines divide the polygon into connect-sum patterns; in this example, the polygon can be decomposed into a start unknot pattern on the left, a proper connect-sum trefoil knot pattern in the middle, and an end unknot pattern on the right.}
\label{fig2tubedef}
\end{figure}

We define the \emph{span} $s(\pi)$ of a polygon $\pi\in\mathcal P_\tube$ to be the maximal $x$-coordinate reached by any of its vertices and we use $|\pi|$ to denote the number of edges in $\pi$. 
See Figure~\ref{fig2tubedef} for a polygon $\pi$ that fits in a $2\times1$ tube with $s(\pi)=6$ and $|\pi|=36$.
Let $p_{\tube,n}(s)$ be the number of polygons in $\mathcal P_{\tube,n}$ with span $s$. To model a force acting parallel to the $x$-axis, we introduce a fugacity conjugate to polygon span which yields a Boltzmann weight, $e^{fs(\pi)}$,  for each polygon $\pi$. Then the ``fixed-edge'' model partition function is given by
\begin{equation}
Z_{\tube,n}(f)  = \sum_{|\pi| = n} e^{fs(\pi)} = \sum_s p_{\tube,n}(s) e^{fs}.
\end{equation}
For this model, the probability of a polygon $\pi \in \mathcal P_{\tube,n}$ is given by
\begin{equation}
{\mathbb{P}}_{n}^{({\rm{ed}}, f)}(\pi)=\frac{e^{fs(\pi)}}{Z_{\tube,n}(f)  }  .
\end{equation}
Thus $f\ll 0$ corresponds to the ``compressed'' regime while $f\gg 0$ corresponds to the ``stretched'' regime.

The (limiting) \emph{free energy per edge} of polygons in $\tube$ is defined as
$$\mathcal F_\tube(f) = \lim_{n\to\infty}\frac1n \log Z_{\tube,n}(f).$$
This is known \cite{Atapour_2009} to exist for all $f$.

For $f=0$, it has been proved that \cite{Soteros_1989, SW1988}
\begin{align}
\mathcal F_\tube(0) = \log\mu_\tube &= \lim_{n\to\infty} n^{-1}\log p_{\tube,n} \nonumber \\
& < \lim_{n\to\infty} n^{-1}\log c_{\tube,n} \nonumber \\
& < \lim_{n\to\infty} n^{-1}\log p_n = \lim_{n\to\infty} n^{-1}\log c_n \equiv \log \mu, \label{eqnmudef}
\end{align}
where $c_n$ is the number of $n$-step self-avoiding walks (SAWs) in $\mathbb Z^3$ starting at the origin and $\mu$ is their growth constant, and $c_{\tube,n}$ is the number of these confined to $\tube$.

A subset of self-avoiding polygons in $\tube$ are \emph{Hamiltonian} polygons: those which occupy every vertex in a $s\times L\times M$ subtube of $\tube$. In addition to being a useful lower bound for general polygons in the $f<0$ compressed regime, these also serve as an idealized model of tightly packed ring polymers \cite{JerniganII-1998}. We define the number of Hamiltonian polygons, $\pham_{\tube,n}$, to be the number of  $n$-edge polygons in $\mathcal{P}_{\tube,n}$ which have span $s$ and occupy every vertex in an $s\times L\times M$ subtube of $\tube$. 
Defining $W_{\tube}=(L+1)(M+1)$ (the number of vertices in an integer plane $x=i\geq 0$ of the tube), we assume without loss of generality that $L\geq M$;
note that $\pham_{\tube,n} = 0$ if $n$ (even) is not a multiple of $W_{\tube}$.
The following limit has been proved to exist \cite{Beaton_2016} (see also \cite{Eng2014MSc}):
\begin{equation}\label{eqn:ham_growthrate1}
\kappaham_\tube = \log\muham_\tube \equiv \lim_{s\to\infty}\frac{1}{(s+1)W_{\tube}} \log \pham_{\tube,(s+1)W_{\tube}}.
\end{equation}
Furthermore, using this, $\mathcal F_\tube(f) $, the free energy per edge, is bounded as follows:
\begin{equation}
\max\{{f/2},(f/W_{\tube})+ \kappaham_\tube\} \leq \mathcal F_\tube(f) \leq  \max\{{f/W_{\tube}},{f/2}\} +\mathcal F_\tube(0),
\label{eqnfreeenergybounds}
\end{equation}
with $\mathcal F_\tube(f)$ asymptotic to the lower bound 
for  $f\to\infty$ for any $\tube$, and for $f\to -\infty$ for small tube sizes (this is conjectured to be true for any $\tube$), see \cite{Beaton_2016}.
More specifically, in \cite{Beaton_2016} it is established that:
\begin{equation}
\lim_{f\to -\infty} \mathcal F_\tube(f) = \betafull_\tube/W_\tube,
\label{fullpatternlimit}
\end{equation}
where $\betafull$ is the exponential growth rate (as span $s\to \infty$) for full $s$-patterns (defined later in Section~\ref{subsection:transfermatrix} and (\ref{betafulldef})).  It was also established that $\betafull_\tube/W_\tube=\kappaham_\tube$ for all tubes such that $5\geq L\geq M\geq 0$.

\subsection{The fixed-span model}\label{ssec:fixed-span}

Here, we will also be interested in the dual  model, called the ``fixed-span'' model, with partition function given by
\[Q_{\tube,s}(g) = \sum_n p_{\tube,n}(s) e^{g n},\]
and where the probability associated with a span $s$ polygon $\pi$ is given by
\begin{equation}
{\mathbb{P}}_{s}^{({\rm{sp}}, g)}(\pi)=\frac{e^{g|\pi|}}{Q_{\tube,s}(g)  }.  
\end{equation}
For this model, when $g \gg 0$  densely packed (in terms of number of edges per span) polygons dominate the partition function, while when $g \ll 0$ polygons with very few edges per span dominate.  
The associated (limiting) {\it free energy per span} exists \cite{Beaton_2016} (see also~\cite{Atapour2008PhD}):
\begin{equation}
\mathcal G_\tube(g) = \lim_{s\to\infty}\frac1s \log Q_{\tube,s}(g).
\end{equation}

For brevity we will introduce quantities analogous to $\kappa_\tube$ and $\mu_\tube$:
\begin{equation}
\chi_\tube = \log\nu_\tube = \mathcal{G}_\tube(0).
\end{equation}
Note also that Hamiltonian polygons can be counted by span. But for these the length is a constant multiple of the span, so we have the simple relation
\begin{equation}\label{eqn:ham_length_vs_span}
\chiham_\tube = \log \nuham_\tube = \lim_{s\to\infty} \frac1s \log \pham_{\tube,(s+1)W_\tube} =  W_\tube \kappaham_\tube.
\end{equation}

Both models correspond to special cases of the grand canonical partition function 
\[G_{\tube}(f,g) = \sum_s \sum_n p_{\tube,n}(s) e^{g n+ f s}\]
and can be studied using transfer matrix methods~\cite{Atapour2008PhD}.  

Hamiltonian polygons can also be studied using transfer matrices \cite{Beaton_2016,Eng2014MSc} and we will also investigate the fixed-span model where polygons are restricted to being Hamiltonian. In that case, the probability associated with a span $s$ Hamiltonian polygon $\pi$ is given by
\begin{equation}
{\mathbb{P}}_{s}^{\mathrm{H}}(\pi)=\frac{1}{\pham_{\tube,(s+1)W_\tube} }  .
\end{equation}

\subsection{Knotting statistics in tubes - theory and exact results}

In terms of knotting statistics for these models, the FWD conjecture has been proved.  For this,  there are known ``pattern theorems'' available for the fixed-edge and fixed-span models  (see~\cite{Atapour2008PhD,Atapour_2009}), as well as for Hamiltonian polygons (see~\cite{Eng2014MSc}).   
The theorems focus on proper polygon patterns (see~\cite{Beaton_2016} for more precise definitions).  
Given a proper pattern $P$ which can occur in a polygon in $\tube$,   define $p_{\tube,n}(s;P,m)$ to be the
number of polygons counted in $p_{\tube,n}(s)$  which contain at most $m$ translates of $P$.      Then we have that, for any fixed $f$ and proper polygon pattern $P$,  there exists $\epsilon_P>0$ such that:
\begin{align}
\mathcal F_\tube(f;P,\epsilon_P) 
& \equiv \limsup_{n\to\infty}\frac1n \log Z_{\tube,n}(f;P,\epsilon_P) \nonumber \\
& \equiv \limsup_{n\to\infty}\frac1n \log \sum_s p_{\tube,n}(s;P,\epsilon_Pn)e^{fs} < \mathcal F_\tube(f). \label{fixededgepatternthm}
\end{align}
So for $n$ sufficiently large, all but exponentially few $n$-edge polygons subjected to the force $f$ contain more than $\epsilon_Pn$ copies of $P$.
Similarly for any fixed $g$ and proper polygon pattern $P$,  there exists $\tilde{\epsilon}_P>0$ such that:
\begin{align}
\mathcal G_\tube(g;P,\tilde{\epsilon}_P) 
&\equiv \limsup_{s\to\infty}\frac1s \log Q_{\tube,s}(g;P,\tilde{\epsilon}_P) \nonumber \\
&\equiv \limsup_{s\to\infty}\frac1s \log \sum_n p_{\tube,n}(s;P,\tilde{\epsilon}_Ps)e^{gn} < \mathcal G_\tube(g). \label{fixedspanpatternthem}
\end{align}
Finally, for any proper Hamiltonian polygon pattern $P$ (see~\cite{Eng2014MSc}), there exists $\hat{\epsilon}_P>0$ such that:
\begin{equation}
\limsup_{n\to\infty} \pham_{\tube,n}(P,\hat{\epsilon}_Pn)< \kappaham_\tube . \label{Hamiltonianpatternthm}
\end{equation}
So all but exponentially few sufficiently long $n$-edge Hamiltonian polygons contain more than $\hat{\epsilon}_P n$ copies of $P$.

In a previous study \cite{BeatonSoftMatter_2018} we defined {\it connect-sum} knot patterns for polygons in tubes. This concept will be useful in the present article, so we briefly review the definition here. For $k\in\mathbb N$, we say a polygon $\pi$ in $\tube$ has a \emph{2-section} at $k+\frac12$ if $\pi$ intersects the plane $x=k+\frac12$ at exactly two points. If $\pi$ has $t$ 2-sections, then it can be partitioned into $t+1$ pieces -- these are \emph{connect-sum patterns} (so named because they give an easy way of writing the polygon as the connect-sum of smaller pieces). If $t\geq 1$ then the first and last are \emph{start} and \emph{end} connect-sum patterns; the remainder (if any) are \emph{proper} connect-sum patterns. (Figure \ref{fig2tubedef} shows a polygon divided into a start, proper and end connect-sum pattern.) Any connect-sum pattern can be converted into a closed curve by joining any pairs of loose ends at the left and at the right; note that if the resulting closed curve has knot-type $K$ then $K$ must be part of the knot decomposition of the original polygon.

\begin{figure}[h]
\centering
\resizebox{0.6\textwidth}{!}{
\includegraphics[width=\textwidth]{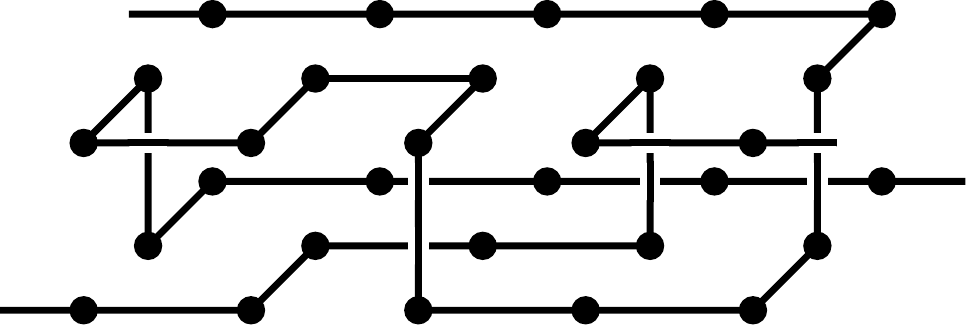}
}
\caption{A minimal-size Hamiltonian connect-sum trefoil pattern in the $2\times1$ tube. This is one of the 32 counted in the first row of Table~\ref{table:jeremy_data}.}
\label{fig:ham_trefoil_pattern}
\end{figure}

Figure~\ref{fig:ham_trefoil_pattern} shows an example of a smallest connect-sum trefoil knot pattern that can occur in a Hamiltonian polygon in a $2\times 1$ tube.    Similar knot patterns can be found for any tube size, so that the pattern theorems of (\ref{fixededgepatternthm})-(\ref{Hamiltonianpatternthm}) can be used to establish that all but exponentially few sufficiently long polygons (regardless of the model and the fixed value of $f$ or $g$), are knotted. 
From such theorems it can also be shown (using arguments analogous to those in \cite{SSW}) that the knot-complexity of polygons grows as polygon ``size'' grows (size could be measured in terms of edges or span), so that a typical polygon will have a highly-composite knot-type $K=K_1\# K_2 \# \ldots\# K_r$. 
Furthermore, for each model, if polygons are restricted to being unknots, the resulting limiting free energy exists and is strictly less than the corresponding limiting free energy for all polygons in the model.  In particular, respectively for the $f=0$ fixed-edge, $g=0$ fixed-span, and the Hamiltonian models, it is known that (see~\cite{Atapour2008PhD,Atapour_2009,Eng2014MSc}) the following limits exist and satisfy:
\begin{equation}\label{eqn:edge_unknotgrowthrate1}
\kappa_{\tube,0_1} = \log\mu_{\tube,0_1} \equiv \lim_{n\to\infty}\frac{1}{n} \log p_{\tube,n}(0_1)< \kappa_{\tube};
\end{equation}
\begin{equation}\label{eqn:span_unknotgrowthrate1}
\chi_{\tube,0_1} = \log\nu_{\tube,0_1} \equiv \lim_{s\to\infty}\frac{1}{s} \log q_{\tube,s}(0_1)< \chi_{\tube},
\end{equation}
where $q_{\tube,s}(K)$ is the number of knot-type $K$ span-$s$ polygons in $\tube$ counted up to $x$-translation; and 
\begin{equation}\label{eqn:ham_unknotgrowthrate1}
\kappaham_{\tube,0_1} = \log\muham_{\tube,0_1} \equiv \lim_{s\to\infty}\frac{1}{(s+1)W_{\tube}} \log \pham_{\tube,(s+1)W_{\tube}}(0_1)< \kappaham_{\tube},
\end{equation}
where $\pham_{\tube,n}(K)$ is the number of $n$-edge knot-type $K$ Hamiltonian polygons in $\tube$ counted up to $x$-translation. We also have $\chiham_{\tube,0_1}\equiv \log \nuham_{\tube,0_1}\equiv W_{\tube}\kappaham_{\tube,0_1}$.




For  small tube sizes ($2\times1$ and $3\times1$ tubes), we previously \cite{BeatonSoftMatter_2018} used  exact generation  to determine all smallest-span connect-sum trefoil knot patterns.
(See Figures~\ref{fig2tubedef} and \ref{fig:ham_trefoil_pattern} for such patterns in the $2\times 1$ tube.) 
Counts are shown in Table~\ref{table:jeremy_data}.

\begin{figure}
\centering
\resizebox{0.75\textwidth}{!}{
\includegraphics[width=\textwidth]{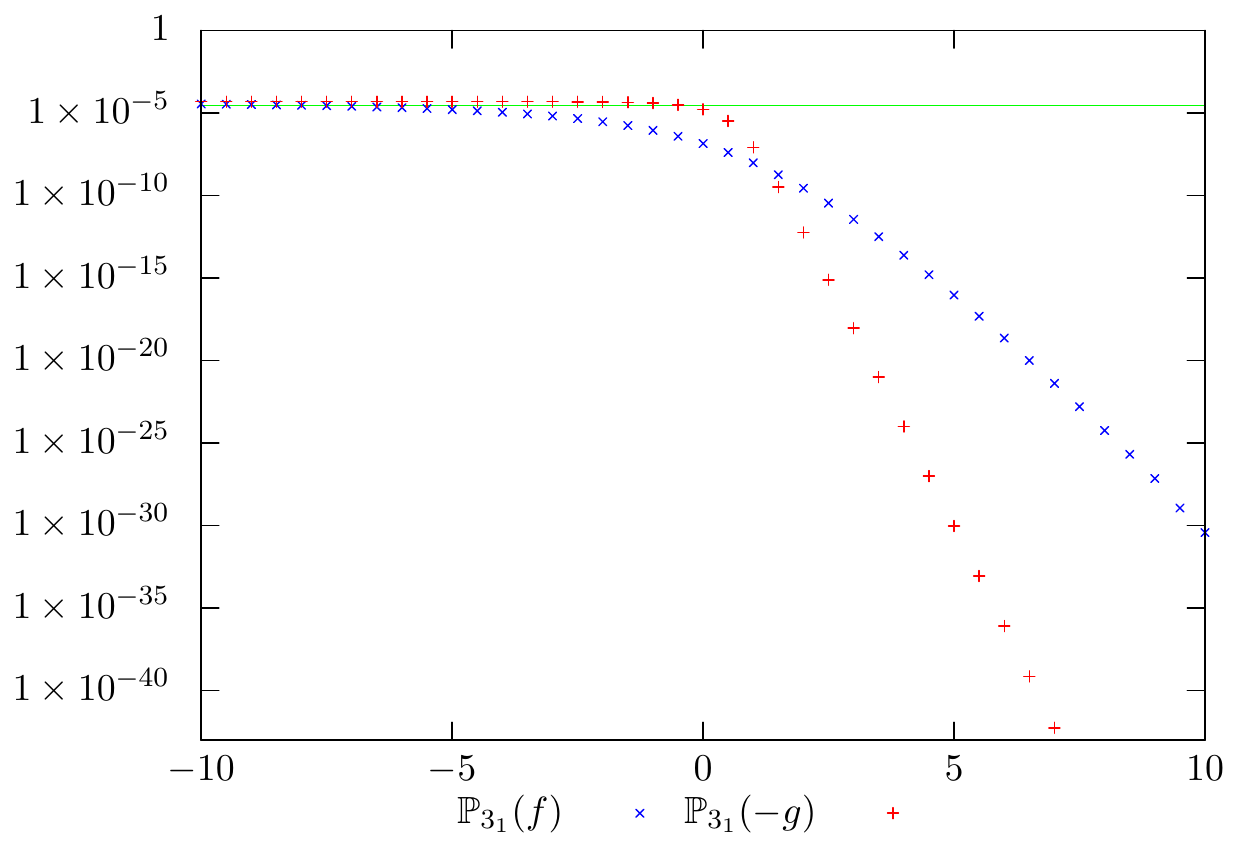}
}
\caption{Log scale plot of the probabilities of the smallest  trefoil patterns in the $3\times1$ tube, as functions of $f$ (blue) and $-g$ (red).  }
\label{fig_fg}
\end{figure}

Towards exploring how the knot statistics depend on the model used (fixed-edge or fixed-span), limiting probabilities of occurrence of these smallest trefoil patterns were determined under each of the distributions ${\mathbb{P}}_{n}^{({\rm{ed}}, f)}$ ($-\infty <f <\infty$, $n\to\infty$) and ${\mathbb{P}}_{s}^{({\rm{sp}}, g)}$ ($-\infty <g<\infty$, $s\to\infty$).  These limiting probabilities can be determined (see 
Lemma \ref{lem:transition_probs_length} in Section~\ref{subsec:montecarlo} and more generally  
\cite{Eng2014MSc}) 
from the eigenvalues and eigenvectors of the transfer matrix.  Figure~\ref{fig_fg} shows the results for the $3\times 1$ tube.  In this figure, 
for the fixed-edge model (${\mathbb{P}}_{n}^{({\rm{ed}}, f)}$),  $\mathbb{P}_{3_1}^{\rm{ed}}(f)$ denotes the limiting ($n\to\infty$) probability of occurrence of a smallest trefoil knot pattern at a section of a polygon. 
Similarly, for the fixed-span model (${\mathbb{P}}_{s}^{({\rm{sp}}, g)}$),  $\mathbb{P}_{3_1}^{\rm{sp}}(g)$ denotes the limiting ($s\to\infty$) probability of occurrence of a smallest trefoil knot pattern at a section of a polygon. 
Further note that Figure~\ref{fig_fg} shows the results for the fixed-edge probabilities with the horizontal axis corresponding to  $f$ while for the fixed-span probabilities it corresponds to $-g$.  
The latter was done to make an easier comparison between the models, since positive values of $f$ and negative values of $g$ both have a stretching effect on polygons.
Although not shown here, the observed trends were similar for the $2\times 1$ tube. 
We observe that, for $\tube_{2,1}$ and $\tube_{3,1}$,  the limiting occurrence probability
of the smallest-span  trefoil knot patterns  decreases (resp.~increases) monotonically with $f$ (resp.~$g$) and approaches a value slightly above the Hamiltonian polygon occurrence probability as $f\to -\infty$ (resp.~$g\to\infty$).     In Section~\ref{subsection:transfermatrix} we explain that this is due to the existence of ``full" trefoil knot patterns which are non-Hamiltonian.

In addition to determining smallest span trefoil patterns, exact counts were obtained for $q_{\tube,s}(K)$ and $\pham_{\tube,(s+1)W_{\tube}}(K)$ for some small spans $s$ and for four different tube sizes. The method used is outlined in \cite{Eng2014MSc} and is based on information gained from the relevant transfer matrices.  The knot-types of all polygons were determined and the resulting counts are shown in Tables \ref{tablePolyGenResults} and \ref{tableHamPolyGenResults}.
Counts shown for $q_{\tube,s}(K)$ have been further delineated by the number of edges $n$ and such counts are available from the authors by request.  
Note that total counts for Hamiltonian polygons in $2\times 1$, $3\times 1$ and $2\times 2$ tubes were published in 1998 \cite[TABLE III]{JerniganII-1998} and our total counts for these tubes confirm the 1998 results; counts by knot-type were not considered in \cite{JerniganII-1998}.

\begin{table}
\caption{Numbers of trefoil patterns of smallest spans in the $2\times1$ and $3\times1$ tubes, for all and Hamiltonian polygons}
\centering
\begin{tabular}{crrrrr} \hline
	Tube & & & & Ham. & Full \\
	Size & Span & Count &  & Count &  Count\\
	\hline\hline
	$2\times1$		& 5 & 116 		& 	& 32		& 36 \\
				& 6 & 5,888	& 	& 748	& 788 \\
				& 7 & 156,224	& 	& 9,408	& 9,928 \\
	\hline
	$3\times1$		& 3 & 1,964	&	& 232	& 276 \\
				& 4 & 792,256	&	& 19,016	& 22,888 \\
	\hline
\end{tabular}
\label{table:jeremy_data}
\end{table}

\begin{table}
\caption{SAP generation results.}
\label{tablePolyGenResults}
\centering
\begin{tabular}{|c|c|r|r|r|r|}
	\hline
	Tube & Span & \multicolumn{1}{c}{Total}\vline & \multicolumn{1}{c}{$3_1^+$}\vline & \multicolumn{1}{c}{$3_1^-$}\vline & \multicolumn{1}{c}{$4_1$}\vline\\
	\hline\hline
	$2\times1$ & 1 & 219				& 0		& 0		& 0		\\
	$2\times1$ & 2 & 7,631			& 0		& 0		& 0		\\
	$2\times1$ & 3 & 264,543			& 0		& 0		& 0		\\
	$2\times1$ & 4 & 9,101,347		& 0		& 0		& 0		\\
	$2\times1$ & 5 & 312,733,719		& 0		& 0		& 0		\\
	$2\times1$ & 6 & 10,745,324,481	& 1,832	& 1,832	& 0		\\
	\hline
	$3\times1$ & 1 & 1,528			& 0		& 0		& 0		\\
	$3\times1$ & 2 & 277,400			& 0		& 0		& 0		\\
	$3\times1$ & 3 & 47,368,928		& 598	& 598	& 0		\\
	$3\times1$ & 4 & 7,863,265,372		& 382,257	& 382,257	& 36		\\
	\hline
	$4\times1$ & 1 & 10,197			& 0		& 0		& 0		\\
	$4\times1$ & 2 & 9,633,793		& 0		& 0		& 0		\\
	$4\times1$ & 3 & 7,939,543,353		& 383,543	& 383,543	& 36		\\
	\hline
	$2\times2$ & 1 & 8,052			& 0		& 0		& 0		\\
	$2\times2$ & 2 & 3,410,348		& 0		& 0		& 0		\\
	$2\times2$ & 3 & 1,430,358,664		& 4,182	& 4,182	& 0		\\
	\hline
\end{tabular}
\end{table}

\begin{table}
\caption{Hamiltonian SAP generation results.}
\label{tableHamPolyGenResults}
\centering
\resizebox{\columnwidth}{!}{
\begin{tabular}{|c|c|r|r|r|r|r|r|r|r|r|r|r|r|r|r|}
	\hline
	Tube & Span & \multicolumn{1}{c}{Total}\vline & \multicolumn{1}{c}{$3_1^+$}\vline & \multicolumn{1}{c}{$3_1^-$}\vline & \multicolumn{1}{c}{$4_1$}\vline & \multicolumn{1}{c}{$5_1^+$}\vline & \multicolumn{1}{c}{$5_1^-$}\vline & \multicolumn{1}{c}{$5_2^+$}\vline & \multicolumn{1}{c}{$5_2^-$}\vline & \multicolumn{1}{c}{$6_1^+$}\vline & \multicolumn{1}{c}{$6_1^-$}\vline & \multicolumn{1}{c}{$3_1^+ \# 3_1^-$}\vline & \multicolumn{1}{c}{$8_{19}$}\vline & \multicolumn{1}{c}{$8_{19}^-$}\vline \\
	\hline\hline
	$2\times1$ & 1 & 22			& 0		& 0		& 0		& 0	& 0	& 0	& 0	& 0	& 0	& 0	& 0	& 0\\
	$2\times1$ & 2 & 324			& 0		& 0		& 0		& 0	& 0	& 0	& 0	& 0	& 0	& 0	& 0	& 0\\
	$2\times1$ & 3 & 4,580		& 0		& 0		& 0		& 0	& 0	& 0	& 0	& 0	& 0	& 0	& 0	& 0\\
	$2\times1$ & 4 & 64,558		& 0		& 0		& 0		& 0	& 0	& 0	& 0	& 0	& 0	& 0	& 0	& 0\\
	$2\times1$ & 5 & 908,452		& 0		& 0		& 0		& 0	& 0	& 0	& 0	& 0	& 0	& 0	& 0	& 0\\
	$2\times1$ & 6 & 12,788,368	& 144	& 144	& 0		& 0	& 0	& 0	& 0	& 0	& 0	& 0	& 0	& 0\\
	$2\times1$ & 7 & 180,011,762	& 4,302	& 4,302	& 0		& 0	& 0	& 0	& 0	& 0	& 0	& 0	& 0	& 0\\
	$2\times1$ & 8 & 2,533,935,102& 96,620	& 96,620	& 72		& 0	& 0	& 0	& 0	& 0	& 0	& 0	& 0	& 0\\
	\hline
	$3\times1$ & 1 & 82			& 0		& 0		& 0		& 0	& 0	& 0	& 0	& 0	& 0	& 0	& 0	& 0\\
	$3\times1$ & 2 & 4,580		& 0		& 0		& 0		& 0	& 0	& 0	& 0	& 0	& 0	& 0	& 0	& 0\\
	$3\times1$ & 3 & 232,908		& 58		& 58		& 0		& 0	& 0	& 0	& 0	& 0	& 0	& 0	& 0	& 0\\
	$3\times1$ & 4 & 11,636,834	& 5,710	& 5,710	& 16		& 0	& 0	& 0	& 0	& 0	& 0	& 0	& 0	& 0\\
	$3\times1$ & 5 & 578,377,118	& 458,980	& 458,980	& 3,216	& 32	& 32	& 70	& 70	& 2	& 2	& 36	& 0	& 0\\
	\hline
	$4\times1$ & 1 & 306			& 0			& 0			& 0		& 0		& 0		& 0	& 0	& 0	& 0	& 0	& 0	& 0\\
	$4\times1$ & 2 & 64,558		& 0			& 0			& 0		& 0		& 0		& 0	& 0	& 0	& 0	& 0	& 0	& 0\\
	$4\times1$ & 3 & 11,636,834	& 5,710		& 5,710		& 16		& 0		& 0		& 0	& 0	& 0	& 0	& 0	& 0	& 0\\
	$4\times1$ & 4 & 2,040,327,632	& 2,264,820	& 2,264,820	& 35,816	& 3,148	& 3,148	& 8	& 8	& 0	& 0	& 0	& 4	& 4\\
	\hline
	$2\times2$ & 1 & 324				& 0		& 0		& 0	& 0	& 0	& 0	& 0	& 0	& 0	& 0	& 0	& 0\\
	$2\times2$ & 2 & 0				& 0		& 0		& 0	& 0	& 0	& 0	& 0	& 0	& 0	& 0	& 0	& 0\\
	$2\times2$ & 3 & 3,918,744		& 96		& 96		& 0	& 0	& 0	& 0	& 0	& 0	& 0	& 0	& 0	& 0\\
	$2\times2$ & 4 & 0				& 0		& 0		& 0	& 0	& 0	& 0	& 0	& 0	& 0	& 0	& 0	& 0\\
	\hline
\end{tabular}
}
\end{table}

Determining the knot-type of a polygon or a knot pattern requires the whole polygon or knot pattern.  However, since the numbers of polygons and knot patterns in a tube grow exponentially with either span or the number of edges,   exact generation has so far been limited to the cases shown in Tables \ref{table:jeremy_data}, \ref{tablePolyGenResults} and \ref{tableHamPolyGenResults}.  To explore knotting statistics further,  a Monte Carlo approach was developed to generate random polygons in the tube, based on a method of \cite{Alm_1990}. The Monte Carlo method is also based on transfer-matrices and can be used to generate a set of independent and identically distributed polygons from any of the distributions
$\left\{{\mathbb{P}}_{n}^{({\rm{ed}}, f)},{\mathbb{P}}_{s}^{({\rm{sp}}, g)}\right\}$ provided that the transfer matrix associated with  $G_{\tube}(f,g)$ is known. 
 Details of the approach are given in Section~\ref{subsec:montecarlo}.  
Based on the exact results of Figure \ref{fig_fg},  we focused primarily on the Hamiltonian polygon model
where the probabilities of the 
smallest trefoil patterns were large (compared to the other models) and where the knot probabilities for small spans are also large. 
Similarly we focus on the $3\times1$ tube, since knots are far more common than in $2\times1$ while the transfer matrices are small enough as to make simulations and enumerations reasonably efficient.  

\subsection{Fixed-span vs. fixed-edge}

Here we briefly make a comment about why we are focusing on the fixed-span and Hamiltonian models, instead of the fixed-edge model. Superficially, it is simply because non-trivial knots are far more common in the first two models. For example, in the $3\times1$ tube, we can compare the growth rates of unknots to all polygons:
\begin{align}
\log\left(\frac{\mu_{\tube,0_1}}{\mu_\tube}\right) &\approx -1.2 \times 10^{-7} & & \text{(fixed-edge)} \\
\frac{1}{\langle O_\tube\rangle}\log\left(\frac{\nu_{\tube,0_1}}{\nu_\tube}\right) &\approx -2.1 \times 10^{-5} & & \text{(fixed-span)} \\
\frac{1}{W_\tube}\log\left(\frac{\nuham_{\tube,0_1}}{\nuham_\tube}\right) &\approx -8.9\times10^{-5} & & \text{(Hamiltonian)}
\end{align}
where ${\langle O_\tube\rangle}$ is the average number of occupied vertices per span in the fixed-span model (see Section~\ref{ssec:growth_constants}, (\ref{eqn:Ham_knotprob_comparison}) and (\ref{eqn:fixedspan_knotprob_comparison}) for further explanation of the latter two quantities). We see here that unknots are far less dominant in the fixed-span and Hamiltonian ensembles than in the fixed-edge ensemble.

Of course, this begs the question: why are knots more likely in one ensemble than another? This is because fixed-span polygons tend to be \emph{more dense} than fixed-edge polygons, that is, the average number of edges (or equivalently, vertices) per unit span is greater for fixed-span. For example, again in the $3\times1$ tube, the average density of edges for a long fixed-span polygon is approximately $6.52$, while for a fixed-edge polygon it is only $4.11$. (Of course for a Hamiltonian polygon, it is exactly 8.) A greater density provides more opportunities for strands to get tangled, and thus leads to a higher knot probability.

Regardless of the model, we expect that conjectures analogous to those of  (\ref{pnKtubeinequality}) and (\ref{pnunknottube}) will hold and we focus on the fixed-span models for which we have been able to obtain the most  data.   In that regard, when $K$ is a knot-type, the notation $\mathbb{P}^{(\mathrm{sp})}_s(K)$ ($\mathbb{P}^{\mathrm{H}}_s(K)$) will denote the probability of a polygon having knot-type $K$ according to the fixed-span distribution $\mathbb{P}^{(\mathrm{sp},0)}_s$ ($\mathbb{P}^{\mathrm{H}}_s$) defined in Section \ref{ssec:fixed-span}.    Then, for example, the analogue of the conjectured form~\eqref{pnunknottube} for these two fixed-span models is:
there exist constants (independent of $s$) $C^{\mathrm{sp}}_{\tube,0_1}$ and $\nu_{\tube,0_1}$ such that,
\begin{equation}
 q_{\tube,s}(0_1) \sim C^{(\mathrm{sp})}_{\tube,0_1} (\nu_{\tube,0_1})^s, \qquad {\rm as}\quad s \rightarrow \infty ;
 \label{qnunknottube}
 \end{equation}
and for Hamiltonian polygons,   there exist constants (independent of $s$) $C^{\mathrm{H}}_{\tube,0_1}$ and $\nuham_{\tube,0_1}$ such that,
\begin{equation}
 \pham_{\tube,(s+1)W_{\tube}}(0_1) \sim C^{\mathrm{H}}_{\tube,0_1} (\nuham_{\tube,0_1})^s, \qquad {\rm as}\quad s \rightarrow \infty .
 \label{pnhamunknottube}
 \end{equation}


\section{Numerical approaches and transfer matrix methods} \label{sec:numerical}

We first review the transfer matrix method and then sketch the Monte Carlo approach used to randomly sample self-avoiding polygons and Hamiltonian polygons in a tube $\tube$.

\subsection{Transfer matrix method}
\label{subsection:transfermatrix}

We rely on the definitions for start, proper and end  {\it 1-pattern} for polygons as given in \cite{Beaton_2016}; essentially these are sets of edges and vertices that can occur between two consecutive half-integer $x$-planes in a polygon in $\tube$ along with a pair partition which defines how any endpoints in the first half-integer plane are connected up on the left within a polygon.    By dividing up a polygon with span $s$ at each half-integer $x$-plane (starting at $x=-1/2$ and ending at $x=s+1/2$), a polygon can then be thought of as a sequence of ($s+1$) 1-patterns that starts with a start 1-pattern, followed by $s-1$ proper 1-patterns, and then ends with an end 1-pattern.   We then say a given 1-pattern $q$ can follow another 1-pattern $p$ if $q$ can occur immediately to the right of $p$ in some polygon in $\tube$.   Given an ordering of all proper 1-patterns,  we can then define the transfer matrix $\M$ for polygons in $\tube$ as:
\begin{equation}
\M_{ij}(g) = \begin{cases}e^{ g |j|}, &  \textrm{ if the $j^\mathrm{th}$ 1-pattern can follow the $i^\mathrm{th}$ 1-pattern} \\  0, & \textrm{otherwise}, \end{cases}
\end{equation}
where $|j|$ is the number of polygon edges (1/2 edges contribute 1/2)  in the $j^\mathrm{th}$ 1-pattern.  We can similarly define a start (end) transfer matrix which has a non-zero entry $\mathbf{S}_{ij}(g) = e^{ g(|i|+ |j|)}$ ($\mathbf{E}_{ij}(g) = e^{ g |j|}$ ) when the $j^\mathrm{th}$ proper 1-pattern can follow the $i^\mathrm{th}$ start 1-pattern  (when the $j^\mathrm{th}$ end 1-pattern can follow the $i^\mathrm{th}$ proper 1-pattern).

Using these matrices, the grand-canonical partition function introduced in the last section can now be written as:
\begin{align}
G_{\tube}(f,g) &= \sum_{s\geq 0}  \sum_{n\geq 4} p_{\tube,n}(s) e^{g n+ f s} \nonumber \\
&= \left[ \sum_{s=0}^1  \sum_{n\geq 4} p_{\tube,n}(s) e^{g n+ f s} \right] + \sum_{i,j}\left[\sum_{k\geq 0} e^{f(k+2)}\mathbf{S}(g) \M(g)^k\mathbf{E}(g) \right]_{i,j}  \nonumber \\
&= \left[ \sum_{s=0}^1  \sum_{n\geq 4} p_{\tube,n}(s) e^{g n+ f s} \right] + e^{2f} \sum_{i,j}\left[ \mathbf{S}(g) (\mathbf{I}-e^f\M(g))^{-1}\mathbf{E}(g) \right]_{i,j},
\label{transfermatrix}
\end{align}
where the sum over $i,j$ is over all start 1-patterns ($i$) and end 1-patterns ($j$). Note that for a fixed $f$ or $g$,  the radius of convergence of this partition function is determined by the singularities of $\det(\mathbf{I}-e^f\M(g))$ and hence the eigenvalues of $\M(g)$.  This in turn can be used to determine the limiting free energies defined in the last section.  
Specifically, to determine the fixed-edge model limiting free energy for a given force $f$, set $x=e^g$ and let $R^{\rm{ed}}(f)$ denote the radius of convergence of $G_{\tube}(f,g)$ as a power series in $x$, then 
\begin{equation}
\mathcal F_\tube(f) = \lim_{n\to\infty}\frac1n \log Z_{\tube,n}(f)=-\log R^{\rm{ed}}(f).
\end{equation}
Similarly, to determine the fixed-span model limiting free energy for a given value for $g$, set $y=e^f$ and let $R^{\rm{sp}}(g)$ denote the radius of convergence of $G_{\tube}(f,g)$ as a power series in $y$, then 
\begin{equation}
\mathcal G_\tube(g) = \lim_{s\to\infty}\frac1s \log Q_{\tube,s}(g)=-\log R^{\rm{sp}}(g).
\end{equation}

More generally, \cite{Beaton_2016} also define $s$-patterns to be sets of edges and vertices that can occur between two half-integer $x$-planes, $x=k+1/2$ and $x=s+k+1/2$,  in a polygon in $\tube$ along with a pair partition which defines how any endpoints in the first half-integer plane ($x=k+1/2$) are connected up on the left within a polygon.  The span of an $s$-pattern is thus $s$.   Any $s$-pattern which consists of $(s-1)W$ vertices is called {\it full} and if it can also occur in a Hamiltonian polygon then it is called a Hamiltonian pattern.  While every Hamiltonian pattern is necessarily full, not every full pattern is a Hamiltonian pattern.

In general it is not possible to assign a knot type to an $s$-pattern, as there may be many ways to connect up the ``loose ends'' on the left and right. The exception is when an $s$-pattern is also a connect-sum pattern; in that case there is only one way to join the ends. Table \ref{table:jeremy_data} focuses on connect-sum patterns whose knot type is a trefoil -- it shows the number of full connect sum trefoil patterns (last column) along with the number of those which are Hamiltonian for various spans.

By restricting the transfer matrix defined above to  either proper full or proper Hamiltonian 1-patterns, one can obtain transfer matrices $\M^\mathrm{F}$ and $\M^{\rm{H}}$ for full patterns and Hamiltonian polygons, respectively, and from their spectra determine:
\begin{align}
\label{betafulldef}
\betafull_{\tube} &\equiv \lim_{s\to\infty} s^{-1}\log t_{\tube,s}^\mathrm{F}, \\
\kappaham_\tube &\equiv \lim_{s\to\infty}\frac{1}{(s+1)W_{\tube}} \log \pham_{\tube,(s+1)W_{\tube}},
\end{align}
where $t_{\tube,s}^\mathrm{F}$ is the number of full $s$-patterns.   In \cite{Beaton_2016} it was established that $\betafull_\tube/W_\tube=\kappaham_\tube$ for all tubes such that $5\geq L\geq M\geq 0$; this is because Hamiltonian $s$-patterns are the dominant class amongst  full $s$-patterns.  It is also known that for any tube dimensions
$\lim_{f\to -\infty} \mathcal F_\tube(f) = \betafull_\tube/W_\tube$. 
However, because for any finite $f$  one expects that there will always be a non-zero probability that a non-Hamiltonian full $s$-pattern occurs, we expect that the probability of, for example,  full connect-sum trefoil patterns will be greater than that for Hamiltonian connect-sum trefoil patterns for every finite $f>-\infty$.   We see this in Figure~\ref{fig_fg} where the horizontal line corresponds to the probability of the Hamiltonian trefoil patterns occurring in a Hamiltonian polygon.

\subsection{Monte Carlo method}\label{subsec:montecarlo}

Next we briefly sketch the method used to randomly sample self-avoiding polygons and Hamiltonian polygons in a tube $\tube$. The polygons are sampled uniformly at random from either the fixed-length or fixed-span ensembles. (Note that for Hamiltonian polygons, length determines span, so that these two ensembles are equivalent.) Any two samples are independent, so there is no need to account for correlations.

The method is inspired by one proposed in~\cite{Alm_1990}, but we make some modifications here.
The approach will work for any of the three models (fixed-edge, fixed-span or Hamiltonian polygons).
For the fixed-edge model take $f=0$ and $x=e^g$  in (\ref{transfermatrix}) and set $\M \equiv \M(\log(x))$; in this case each $n$-edge polygon in $\tube$ is considered to be equally likely (uniform).
For a given value of $x$ let $\rho(x)$ be the spectral radius of $\M$, and let $x_0$ be the smallest positive real value of $x$ which makes $\rho(x)=1$. The Perron-Frobenius theorem implies that such a value exists, and moreover that at $x=x_0$, 1 will be a simple eigenvalue\footnote{This requires that $\M$ be irreducible, which can be demonstrated by a straightforward concatenation argument for polygons (see for example \cite{Atapour2008PhD}).} of $\M$. Let $\xi$ (resp.~$\eta$) be the right (resp.~left) eigenvector corresponding to that eigenvalue. Then we have the following lemma for the fixed-edge ensemble.

\begin{lem}[Alm and Janson~\cite{Alm_1990}]\label{lem:transition_probs_length}
Let $i$ and $j$ represent proper 1-patterns such that $j$ can follow $i$. Let ${\mathsf{p}}^{\rm{ed}}_{ij}(n)$ be the probability that an occurrence of $i$ in a uniformly random polygon of length $n$ is followed by $j$. Then as $n\to\infty$,
\begin{equation}
{\mathsf{p}}^{\rm{ed}}_{ij}(n) \to {\mathsf{p}}^{\rm{ed}}_{ij}= x_0^{|j|} \frac{\xi_j}{\xi_i}.
\end{equation}
Furthermore, for any polygon pattern $P$ consisting of a sequence of $b$ internal 1-patterns $\pi_1, \ldots, \pi_b$, the probability, $\mathsf{p}^{\rm{ed}}_{P}(n)$, that $P$ occurs at any given section of a  random length-$n$  polygon satisfies (as $n\to\infty$):
\begin{align}
{\mathsf{p}}^{\rm{ed}}_{P}(n) \to {\mathsf{p}}^{\rm{ed}}_{P}&= ( \eta_{\pi_1} \xi_{\pi_1} ) \left( x_0^{|\pi_2|} \frac{\xi_{\pi_2}}{\xi_{\pi_1}} \right)  \left( x_0^{|\pi_3|} \frac{\xi_{\pi_3}}{\xi_{\pi_2}} \right) \cdots \left( x_0^{|\pi_b|} \frac{\xi_{\pi_b}}{\xi_{\pi_{b-1}}} \right) \nonumber \\
&= \eta_{\pi_1} x_0^{|\pi_2| + |\pi_3| + \cdots + |\pi_b|} \xi_{\pi_b},
\end{align}
where $|\pi_i|$ denotes the number of edges of the 1-pattern.
\end{lem}

For the fixed-span ensemble with $g=0$ (the probability of each span $s$ polygon is equally likely), the above result requires only minor modification. Let $\lambda$ be the dominant eigenvalue of $\M(0)$. By the Perron-Frobenius theorem, $\lambda$ is real, positive and simple. Let $\zeta$ be the corresponding right eigenvector.

\begin{lem}\label{lem:transition_probs_span}
Let $i$ and $j$ represent proper 1-patterns such that $j$ can follow $i$. Let ${\mathsf{p}}^{\rm{sp}}_{ij}(s)$ be the probability that an occurrence of $i$ in a uniformly random polygon of span $s$ is followed by $j$. Then as $s\to\infty$,
\begin{equation}\label{eqn:fixedspan_transition_probs}
{\mathsf{p}}^{\rm{sp}}_{ij}(s) \to {\mathsf{p}}^{\rm{sp}}_{ij} = \lambda^{-1}\frac{\zeta_j}{\zeta_i}.
\end{equation}
\end{lem}

We randomly generate polygons by building them one 1-pattern at a time, using Lemmas~\ref{lem:transition_probs_length} and~\ref{lem:transition_probs_span} to inform our choice of transition probabilities. For the fixed-span ensemble with $g=0$, the procedure for generating a polygon $\pi$, comprised of 1-patterns $\pi_0, \pi_1, \ldots, \pi_s$, is as follows.
\begin{enumerate}
\item $\pi_0$ is selected uniformly at random from all $S$ start 1-patterns.
\item With probability $r_1(\pi_0)$ (detailed below), the sample is rejected and we return to step 1. Otherwise, $\pi_1$ is selected from all proper 1-patterns which can follow $\pi_0$ with probability proportional to $\zeta_{\pi_1}$.
\item For $i=2,3,\ldots,s-1$, choose $\pi_i$ with probability ${\mathsf{p}}^{\rm{sp}}_{\pi_{i-1},\pi_i}$.
\item With probability $r_s(\pi_{s-1})$ (detailed below), the sample is rejected and we return to step 1. Otherwise, $\pi_s$ is selected uniformly from all end 1-patterns which can follow $\pi_{s-1}$.
\end{enumerate}
The probabilities $r_1$ and $r_s$ are chosen so as to make the sampling uniformly random. First define
\begin{equation}
t_1(i) = \sum_{\substack{j \text{ proper} \\ j \text{ follows } i}} \!\!\!\!\zeta_j \quad {\rm{and}}\quad t_s(i) = \frac{\#{\textrm{ end 1-patterns following }} i}{\zeta_i}.
\end{equation}
Then
\begin{equation}
r_1(\pi_0) = 1-\frac{t_1(\pi_0)}{\displaystyle \max_{i ~\rm{ start}} \{t_1(i)\}} \quad{\rm{and}}\quad r_s(\pi_{s-1}) = 1 - \frac{t_s(\pi_{s-1})}{\displaystyle \max_{i ~\rm{ proper}}\{t_s(i)\}}.
\end{equation}

Sampling from the fixed-edge ensemble (with $f=0$) works in a similar way, with the main difference being that the procedure is terminated once the polygon length (rather than the span) reaches the desired value. 

\subsection{Determining the knot type}

A standard technique for determining the knot type of a polygon is to take a projection of the polygon and compute a polynomial invariant, like the Alexander or HOMFLY polynomial. While this does not perfectly distinguish between different knot types (some knots have the same polynomials), such ``collisions'' are expected to be rare for the size of polygons studied here (where the most likely prime knots have small minimum crossing number)  and hence they are unlikely to noticeably affect any statistics.

As will be seen in the next section, polygons of span $s$ in narrow tubes are dominated by unknots until $s$ gets into the thousands. However, the projection of such a large polygon (even an unknot) will contain many hundreds or thousands of crossings. This makes the direct computation of any knot polynomial prohibitively difficult.

Instead, we employ two techniques to simplify this procedure. Firstly, we make use of the aforementioned 2-sections. Wherever a polygon of knot type $K$ contains a 2-section, one can cut it into two pieces, close up the loose ends to form two smaller polygons of knot types $K_1$ and $K_2$, and know that $K = K_1 \# K_2$. By repeating this procedure, a long polygon can be split into a sequence of $p$ smaller polygons, with $K=K_1\#K_2\#\dots\#K_p$. Moreover, the pattern theorems for polygons in $\tube$ imply that long polygons have a positive density of 2-sections, so on average $p=O(s)$. Each of these smaller polygons will have far fewer crossings than the original.

Secondly, we make use of the fact that any polygon on the cubic lattice with length less than $24$ is an unknot~\cite{scharein_bounds_2009}. If any of the small polygons obtained during the ``cutting'' procedure have length 22 or smaller, they are unknots and can immediately be discarded. For those pieces of length 24 or greater, we then apply the BFACF algorithm (see \cite{janse_van_rensburg_monte_2009} and references therein) in an attempt to decrease the size without changing the knot type. BFACF is a Monte Carlo procedure that involves making ``local moves'' to transform a polygon; here we are not using it as a Monte Carlo tool, but rather taking advantage of the fact that (a) it does not change the knot type, and (b) it can be modified to preferentially decrease the size of polygons. If the size drops below 24, we can again discard the piece. (Note that while the original polygon was confined to the tube, when applying BFACF we no longer need the tube restriction.)

After cutting up the polygon and shrinking its constituent pieces as much as possible, we are left with a sequence of sub-polygons of lengths $\geq 24$ and indeterminate knot type. These are then loaded into \textsc{KnotPlot}~\cite{knotplot}, which computes knot polynomials and uses these to determine the knot type.

\section{Results}\label{sec:results}

\subsection{Growth constants}\label{ssec:growth_constants}

For the remainder of this paper we will focus on the fixed-span and Hamiltonian models in the tube, and for fixed-span set $g=0$ (i.e.~no stretching or compressing force). We first consider the growth rates of all polygons. As mentioned in the previous section, these can be determined directly from the transfer matrix, if it is known. The results for the $2\times1$ and $3\times1$ tubes are presented in Table~\ref{table:growth_rates}. For all values the error is expected to be confined to the last digit.

\begin{table}[h]
\caption{Growth rates for all and Hamiltonian polygons in the $2\times1$ and $3\times1$ tubes, counted by span.}
\centering
{\small
\begin{tabular}{cllll} \hline
	Tube & & & & \\
	Size & $\chi_\tube$ & $\nu_\tube$ & $\chiham_\tube$ & $\nuham_\tube$ \\
	\hline\hline
	$2\times1$		& 3.53689835537159142 & 34.3601806741352594 		& 2.644502344846937	& 14.07643812777425	 \\
	\hline
	$3\times1$		& 5.10696921077147344 & 165.1690030762774319	& 3.904865602438742	& 49.643407510907971\\
	\hline
\end{tabular}
}
\label{table:growth_rates}
\end{table}

We note here that, since $\nu_\tube$ and $\nuham_\tube$ are eigenvalues of matrices with integer entries, they are algebraic numbers. However, we have no reason to expect their minimal polynomials to be of low degree, and indeed would expect such polynomials to quickly increase in complexity as tube size grows.

\begin{figure}
\centering
\begin{subfigure}{0.49\textwidth}
\includegraphics[width=\textwidth]{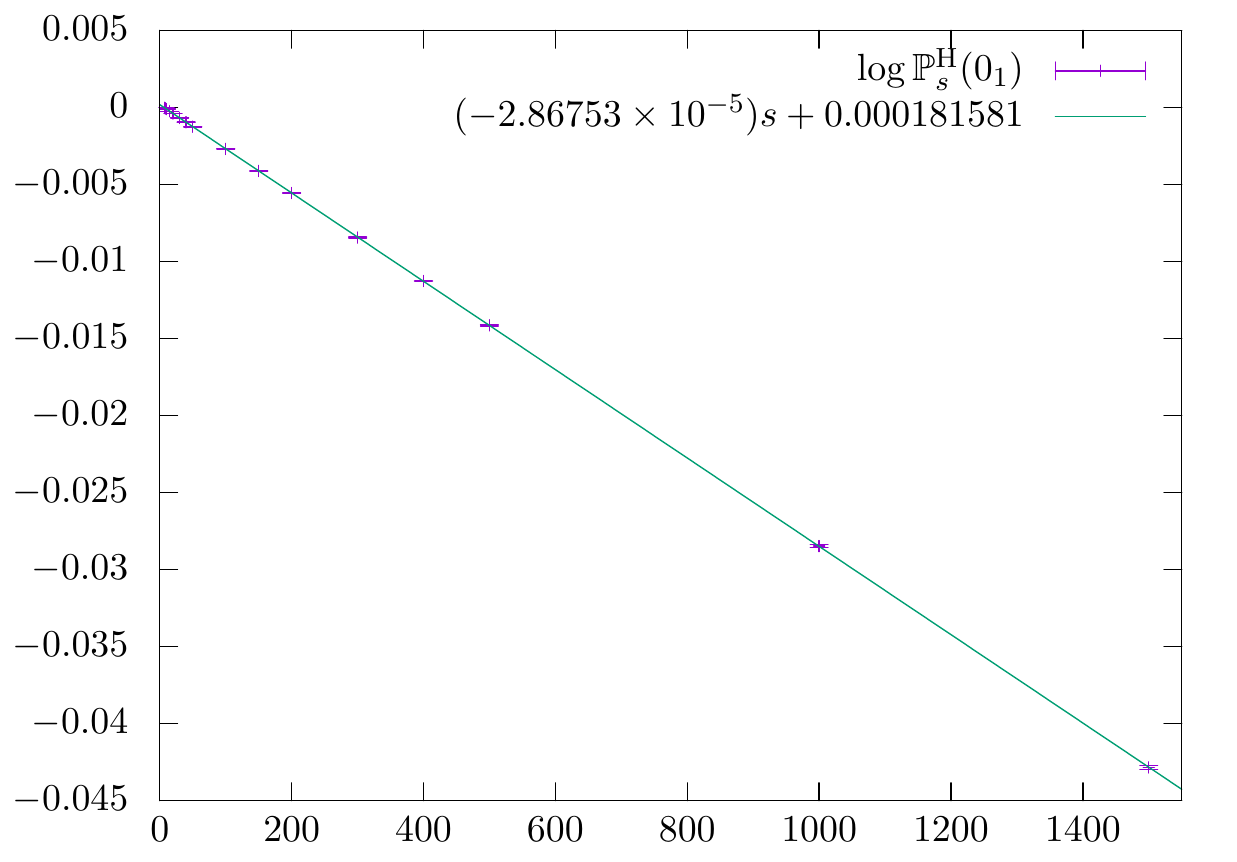}
\caption{}
\label{fig:2x1_Ham_logP0_1_fit}
\end{subfigure}
\begin{subfigure}{0.49\textwidth}
\includegraphics[width=\textwidth]{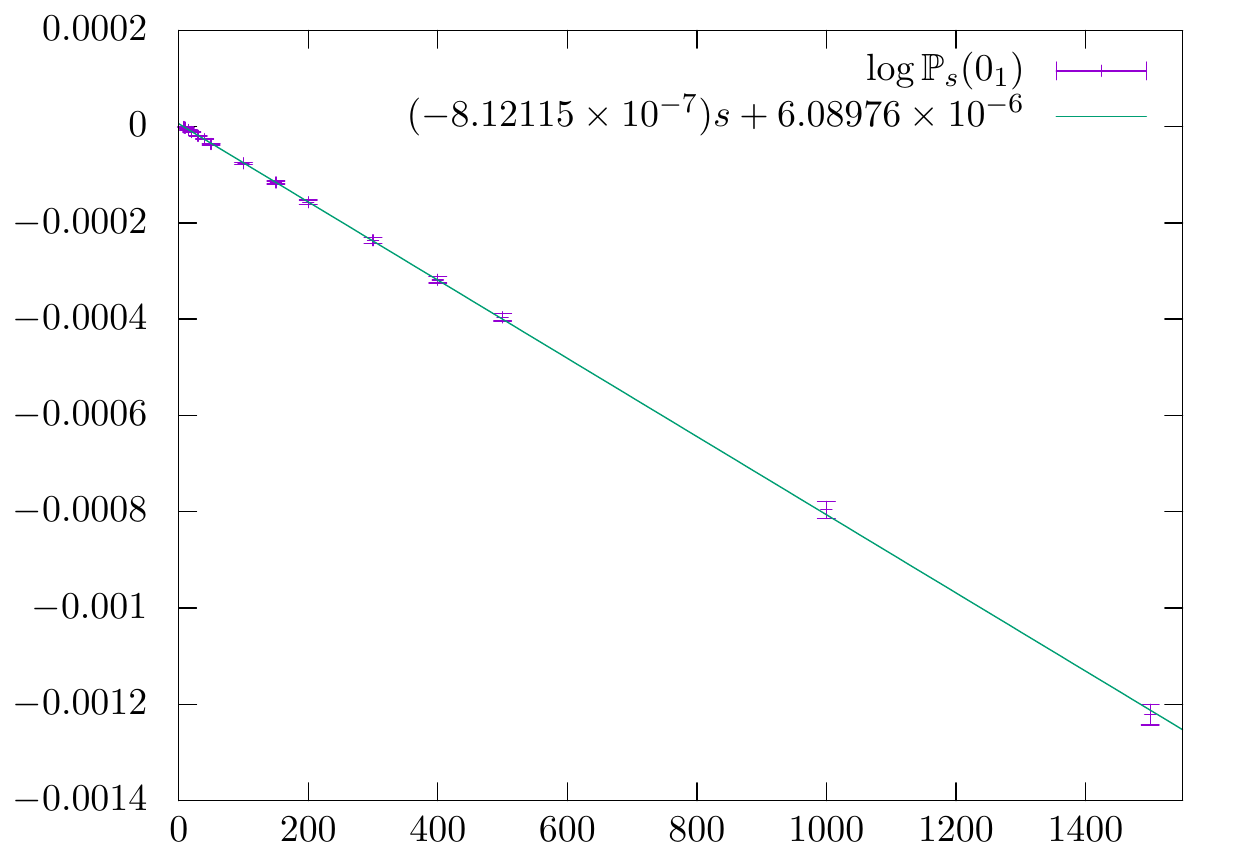}
\caption{}
\label{fig:2x1_all_logP0_1_fit}
\end{subfigure}
\caption{Plots of (a) $\log\mathbb{P}^\mathrm{H}_s(0_1)$ and (b) $\log\mathbb{P}^{(\rm{sp})}_{\tube,s}(0_1)$ against $s$ (span) for the $\tube = 2\times1$ tube, together with a linear best fits.}
\label{fig:2x1-logP0_1-both}
\end{figure}

For Hamiltonian polygons in the $2\times1$ tube, in Figure~\ref{fig:2x1_Ham_logP0_1_fit} we plot $\log \mathbb{P}^\mathrm{H}_s(0_1)$ against $s$ (span), together with a linear best fit.

\begin{figure}
\centering
\begin{subfigure}{0.49\textwidth}
\includegraphics[width=\textwidth]{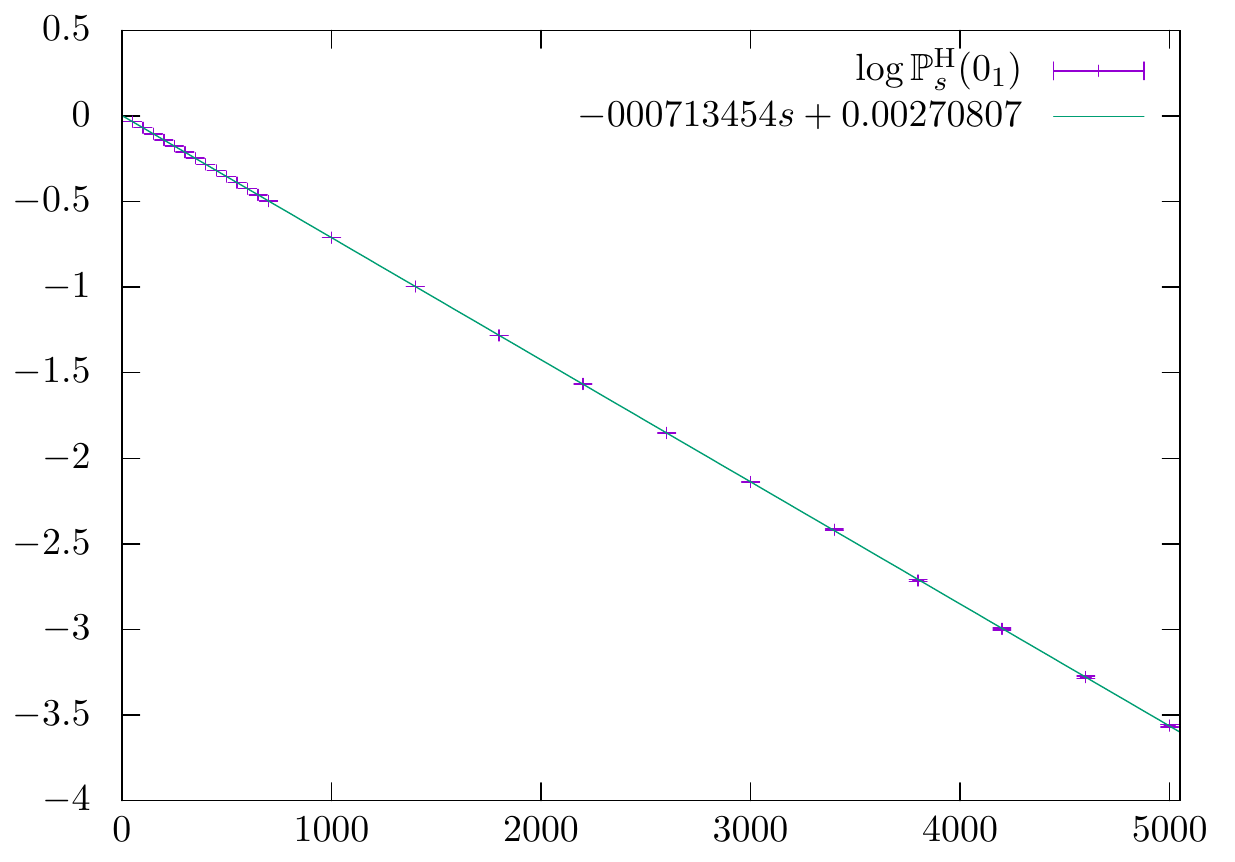}
\caption{}
\label{fig:3x1_Ham_logP0_1_fit}
\end{subfigure}
\begin{subfigure}{0.49\textwidth}
\includegraphics[width=\textwidth]{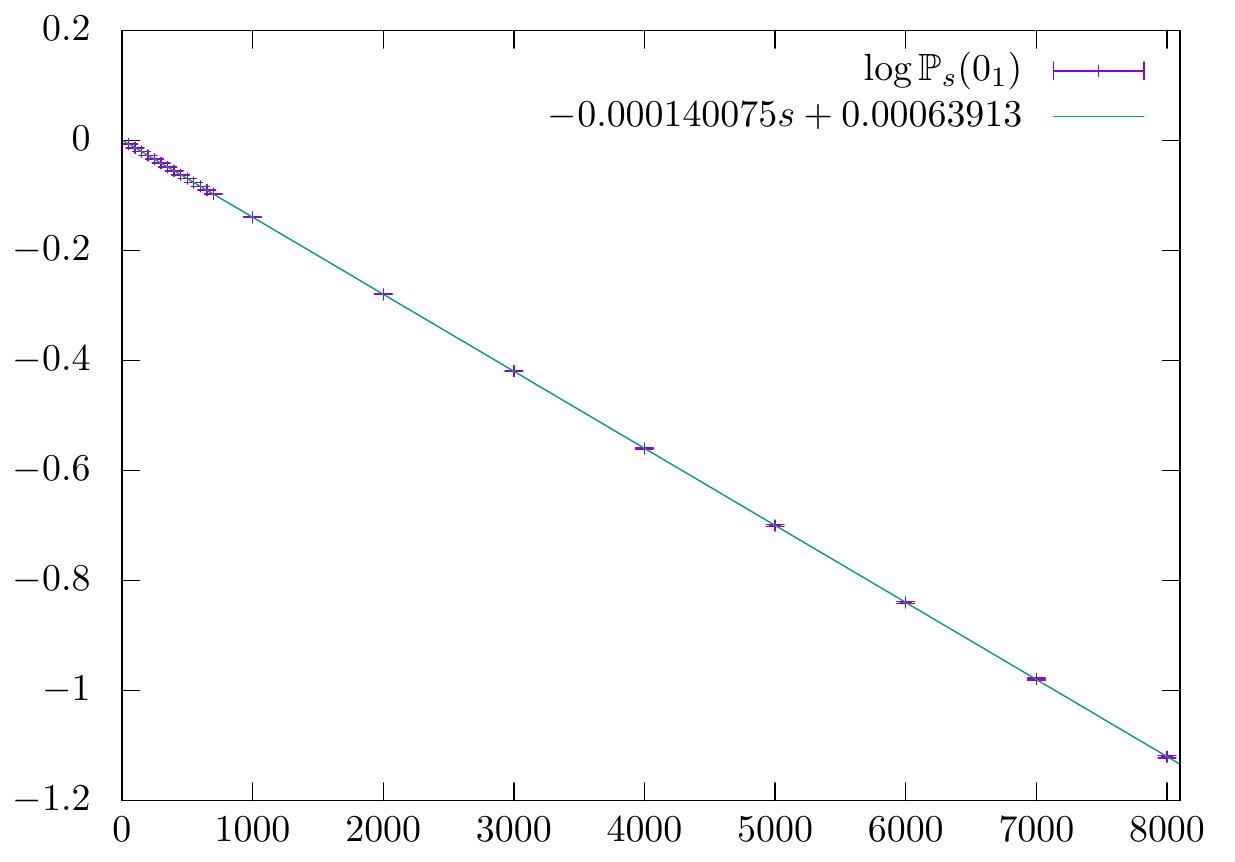}
\caption{}
\label{fig:3x1_all_logP0_1_fit}
\end{subfigure}
\caption{Plots of (a) $\log\mathbb{P}^\mathrm{H}_{\tube,s}(0_1)$ and (b) $\log\mathbb{P}^{(\rm{sp})}_{\tube,s}(0_1)$ against $s$ (span) for the $\tube = 3\times1$ tube, together with a linear best fits.}
\label{fig:3x1-logP0_1-both}
\end{figure}

As is clear from the plot, the linear fit is very good. We therefore can conclude that $\log \mathbb{P}^\mathrm{H}_{\tube,s}(0_1) \sim as+b$, where $a = -2.86753\times10^{-5} \pm 7.105\times10^{-9}$ and $b = 1.81581\times10^{-4} \pm 3.372\times10^{-6}$. Note that there is no need for a power-law correction as per~\eqref{stu-form}. We therefore expect the form
\begin{equation}
\log \mathbb{P}^\mathrm{H}_{\tube,s}(0_1) \sim C + s\log\left(\frac{\nuham_{\tube,0_1}}{\nuham_\tube}\right), \qquad s\to\infty,
\end{equation}
for a constant $C$, and it follows that $\nuham_{\tube,0_1} = 14.0760345$, with errors confined to the last digit.  This provides evidence that the analogue of the conjectured form (\ref{pnunknottube}) for Hamiltonian polygons holds in the $2\times 1$ tube, i.e. that the form (\ref{pnhamunknottube}) is correct.

We repeat this procedure for all polygons in the $2\times1$ tube, as well as Hamiltonian and all polygons in $3\times1$ -- see Figures~\ref{fig:2x1_all_logP0_1_fit}, \ref{fig:3x1_Ham_logP0_1_fit} and \ref{fig:3x1_all_logP0_1_fit}. In each case we have a very good linear fit; providing further evidence that the analogues of the asymptotic form (\ref{pnunknottube}) (namely (\ref{qnunknottube}) and (\ref{pnhamunknottube})) hold. The results for the growth rates of unknotted polygons in all four cases are summarised in Table~\ref{table:unknot_growth_rates}.  

\begin{table}[h]
\caption{Growth rates for all and Hamiltonian unknotted polygons in the $2\times1$ and $3\times1$ tubes, counted by span. Errors are expected to be confined to the last digit.}
\centering
\begin{tabular}{cll} \hline
	Tube & & \\
	Size &  $\nu_{\tube,0_1}$ & $\nuham_{\tube,0_1}$ \\
	\hline\hline
	$2\times1$		& 34.3601527 & 14.0760345	 \\
	\hline
	$3\times1$		& 165.14587 & 49.6080 \\
	\hline
\end{tabular}
\label{table:unknot_growth_rates}
\end{table}

Note that unlike $\nu_\tube$ and $\nuham_\tube$, it is unknown if $\nu_{\tube,0_1}$ or $\nuham_{\tube,0_1}$ are algebraic numbers.

These results should be compared to the best estimates for the unrestricted cubic lattice $\mathbb Z^3$. There, it is estimated~\cite{JansevanRensburg2008} that
\begin{equation}
\log\left(\frac{\mu_{0_1}}{\mu}\right) = (-4.15\pm0.32)\times10^{-6}.
\end{equation}
The figures given in Tables~\ref{table:growth_rates} and~\ref{table:unknot_growth_rates} are for polygons counted by span, not length; however, as previously observed, for Hamiltonian polygons these are in direct proportion. Using~\eqref{eqn:ham_length_vs_span}, we find
\begin{equation}\label{eqn:Ham_knotprob_comparison}
\log\left(\frac{\muham_{\tube,0_1}}{\muham_\tube}\right) = \frac{1}{W_\tube} \log\left(\frac{\nuham_{\tube,0_1}}{\nuham_\tube}\right)\approx \begin{cases} -4.77922 \times 10^{-6} & \tube = 2\times1\text{ tube} \\ -8.91818\times10^{-5} & \tube=3\times1\text{ tube}.\end{cases}
\end{equation}
That is, knots in Hamiltonian polygons in the  $2\times1$ and $3\times1$ tubes are more common than they are in all polygons on the cubic lattice.

For all polygons in the fixed-span ensemble, the lengths of polygons are not fixed and therefore there is no exact way to make a direct comparison with polygons in the full lattice. However, we can use an approximation based on~\eqref{eqn:Ham_knotprob_comparison}, replacing $W_\tube$ -- the number of occupied vertices per unit span for Hamiltonian polygons -- with the corresponding \emph{average} $\langle O_\tube\rangle$ for all polygons. As polygon size gets large, this average can be computed using the stationary distribution of the Markov chain with transition probabilities given by~\eqref{eqn:fixedspan_transition_probs}.

For the $2\times1$ and $3\times1$ tubes, these average vertex densities $\langle O_\tube\rangle$ are 4.8865 and 6.5244 respectively. The analogous version of~\eqref{eqn:Ham_knotprob_comparison} is then
\begin{equation}\label{eqn:fixedspan_knotprob_comparison}
\frac{1}{\langle O_\tube\rangle}\log\left(\frac{\nu_{\tube,0_1}}{\nu_\tube}\right) \approx \begin{cases}
-1.6620\times10^{-7} & \tube=2\times1\text{ tube} \\ -2.1469\times10^{-5} & \tube=3\times1\text{ tube}. \end{cases}
\end{equation}
Using this rough approximation, we see that knots in the $2\times1$ tube are less likely than in the cubic lattice, but more likely in the $3\times1$ tube.

Having established that 
\begin{equation}
\mathbb{P}^{(\mathrm{sp})}_{\tube,s}(0_1) \sim A^{(\mathrm{sp})}_{\tube} \left(\frac{\nu_{\tube,0_1}}{\nu_\tube}\right)^s \qquad\text{and}\qquad\mathbb{P}^\mathrm{H}_{\tube,s}(0_1) \sim A^{\mathrm{H}}_{\tube}\left(\frac{\nuham_{\tube,0_1}}{\nuham_\tube}\right)^s,
\end{equation}
we can see that for $K=0_1$, the asymptotic form~\eqref{probKasymptotics} does indeed appear to hold, with $\Delta N(0_1) \approx 0$. We now wish to investigate if this is still the case with other knot types.   That is, we investigate whether the analogue of the conjectured form~\eqref{pnKtubeinequality} holds.  

\begin{figure}
\centering
\begin{subfigure}{0.49\textwidth}
\includegraphics[width=\textwidth]{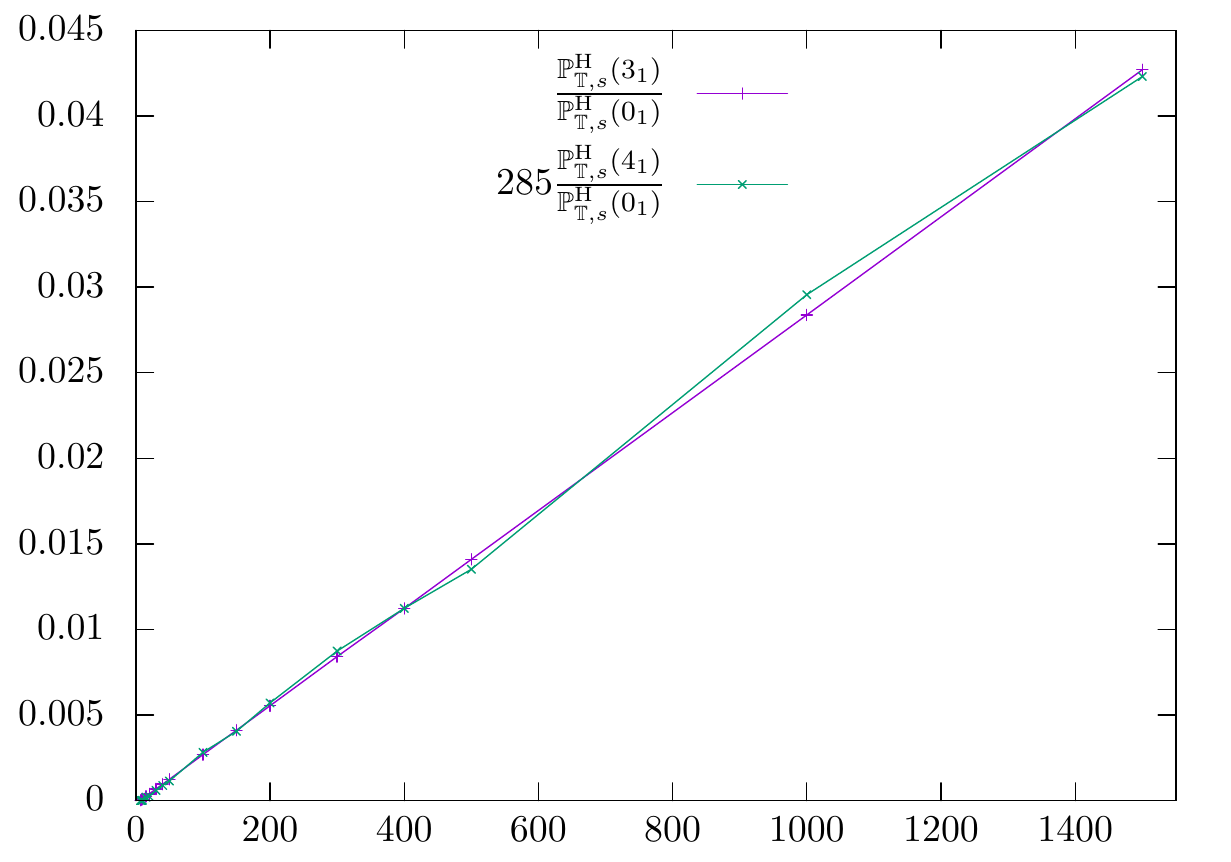}
\caption{}
\label{fig:2x1-Ham-3_1and4_1over0_1}
\end{subfigure}
\begin{subfigure}{0.49\textwidth}
\includegraphics[width=\textwidth]{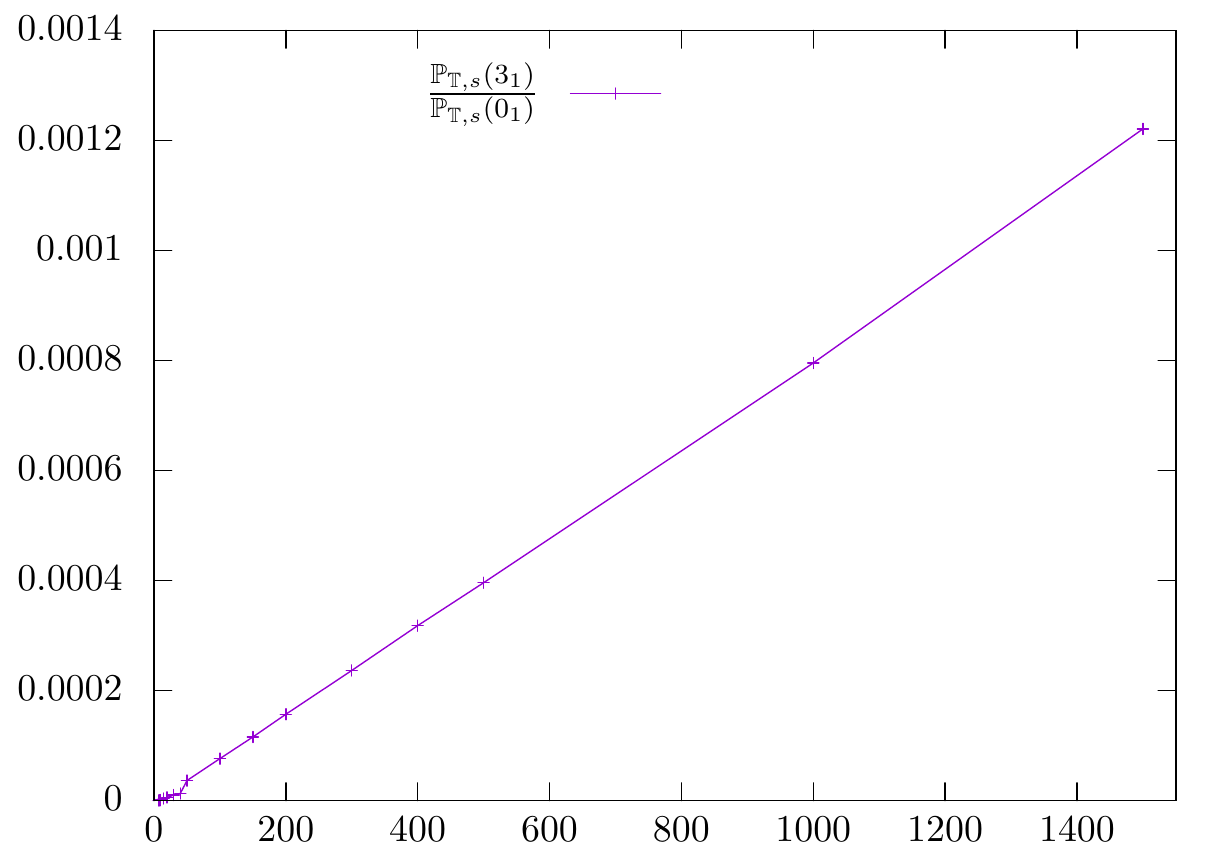}
\caption{}
\label{fig:2x1-all-3_1over0_1}
\end{subfigure}
\caption{(a) A plot of $\mathbb{P}^\mathrm{H}_{\tube,s}(3_1)/\mathbb{P}^\mathrm{H}_{\tube,s}(0_1)$ and $\mathbb{P}^\mathrm{H}_{\tube,s}(4_1)/\mathbb{P}^\mathrm{H}_{\tube,s}(0_1)$ in the $2\times1$ tube, with the latter scaled by a constant factor for clarity. (b) A plot of $\mathbb{P}^{(\mathrm{sp})}_{\tube,s}(3_1)/\mathbb{P}^{(\mathrm{sp})}_{\tube,s}(0_1)$ in the $2\times1$ tube.}
\label{fig:2x1_prime_over_0_1}
\end{figure}

\begin{figure}
\centering
\begin{subfigure}{0.49\textwidth}
\includegraphics[width=\textwidth]{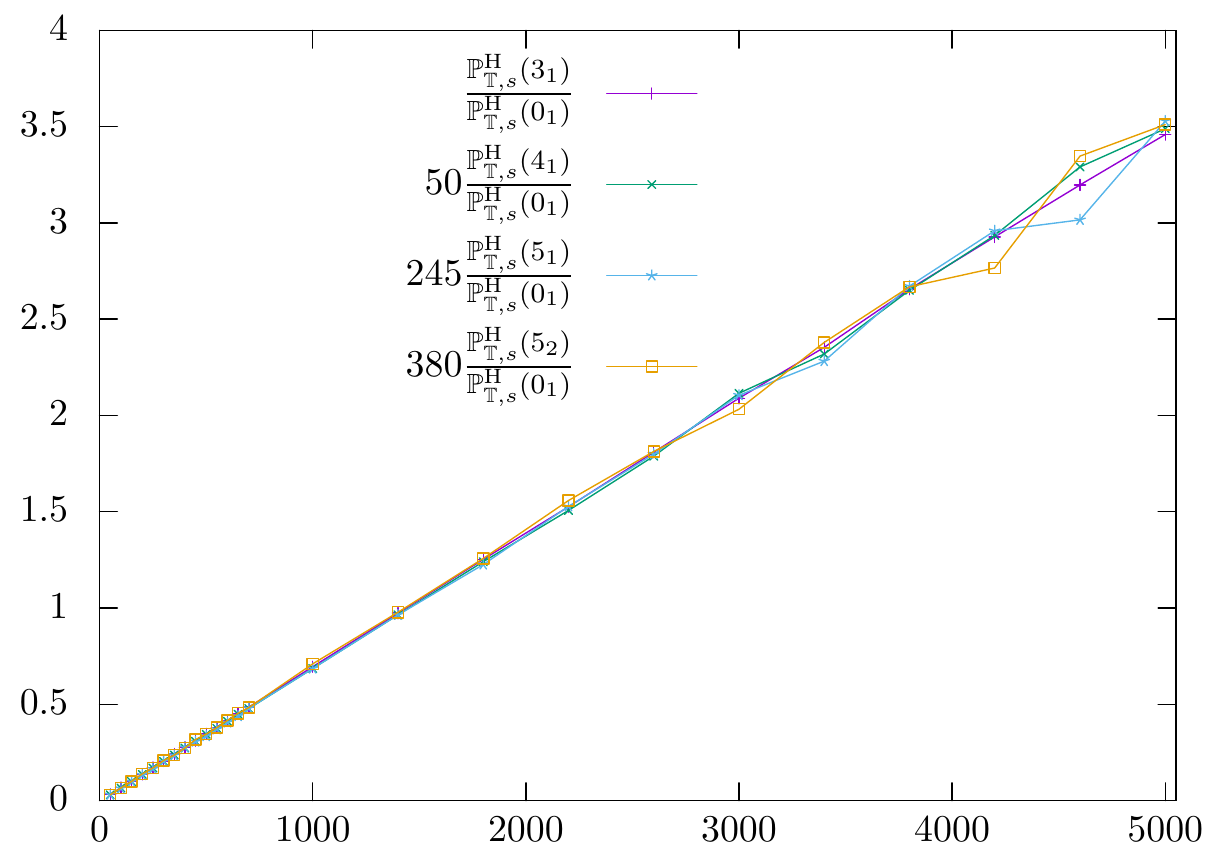}
\caption{}
\label{fig:3x1-Ham-primeover0_1}
\end{subfigure}
\begin{subfigure}{0.49\textwidth}
\includegraphics[width=\textwidth]{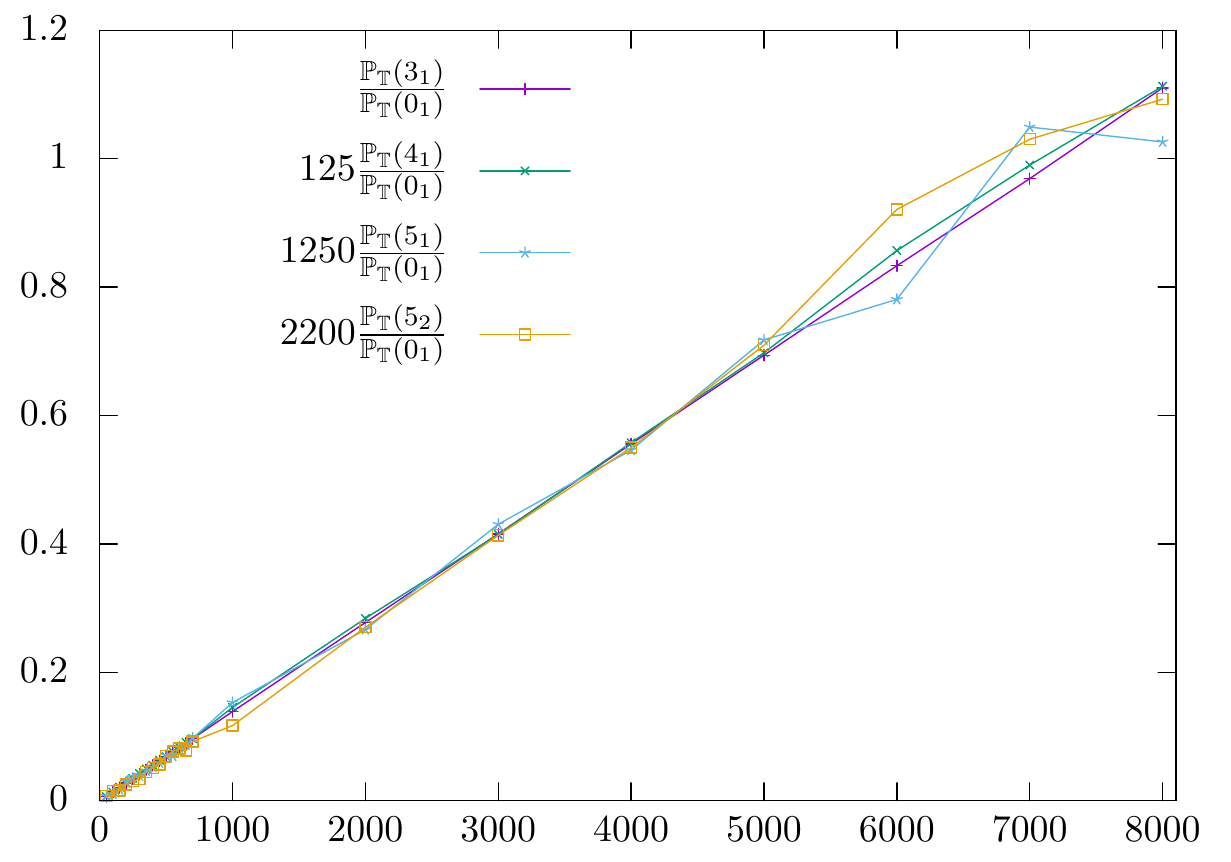}
\caption{}
\label{fig:3x1-all-primeover0_1}
\end{subfigure}
\caption{(a) A plot of $\mathbb{P}^\mathrm{H}_{\tube,s}(K)/\mathbb{P}^\mathrm{H}_{\tube,s}(0_1)$ for $K=3_1,4_1,5_1$ and $5_2$ in the $3\times1$ tube, scaled by constant factors for clarity. (b) The corresponding plots for all polygons in the fixed-span model.}
\label{fig:3x1_prime_over_0_1}
\end{figure}

In Figures~\ref{fig:2x1_prime_over_0_1} and~\ref{fig:3x1_prime_over_0_1} we plot the probabilities of various prime knot types divided by the probability of the unknot, scaled by constant factors so as to be visible in the same plot. The very clear linear form of all these plots (except for some numerical uncertainty at large lengths for the 5-crossing knots) confirms that~\eqref{probKasymptotics} remains applicable for different prime knot types, with $\Delta N(K) \approx 0$ in all cases.

\begin{figure}
\centering
\begin{subfigure}{0.49\textwidth}
\includegraphics[width=\textwidth]{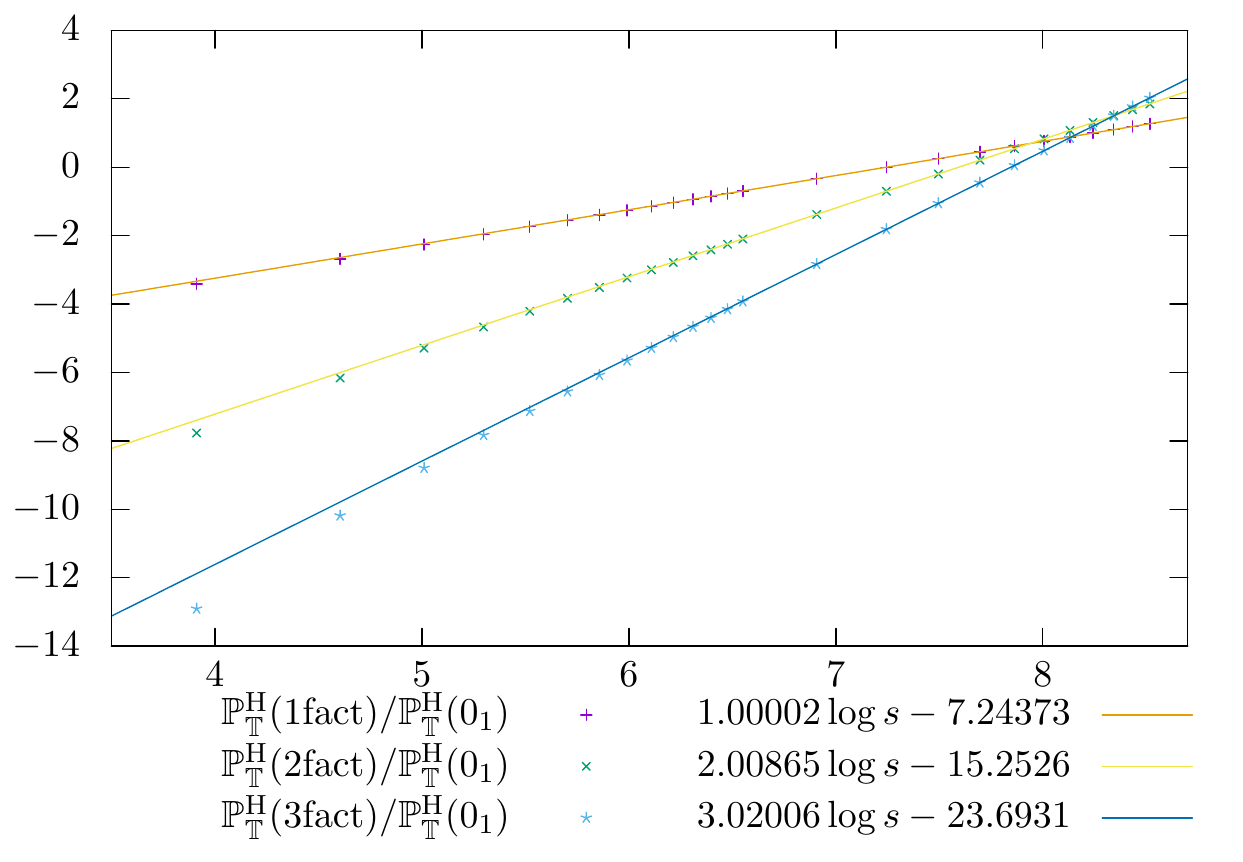}
\caption{}
\label{fig:3x1-Ham-1and2and3_components_loglogfit}
\end{subfigure}
\begin{subfigure}{0.49\textwidth}
\includegraphics[width=\textwidth]{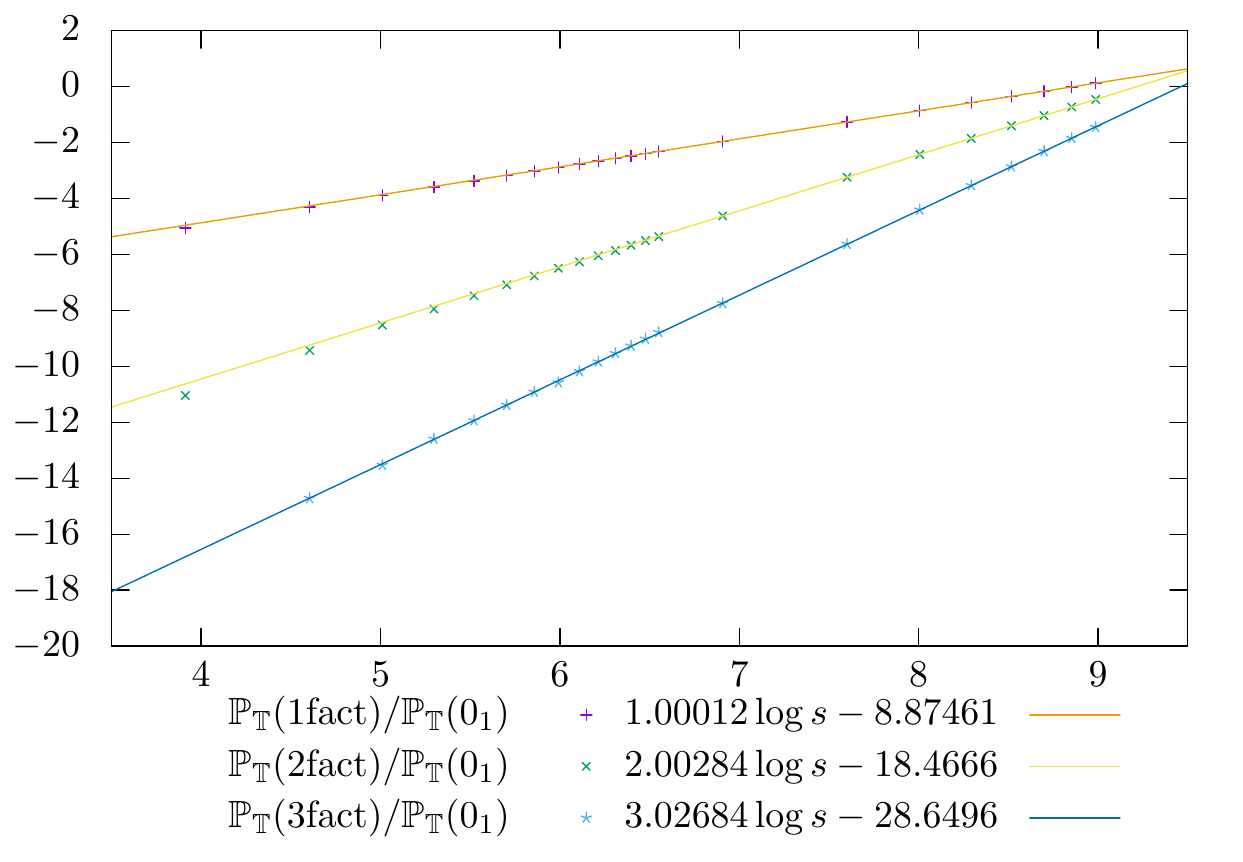}
\caption{}
\label{fig:3x1-all-1and2and3_components_loglogfit}
\end{subfigure}
\caption{(a) A log-log plot of $\mathbb{P}^\mathrm{H}_{\tube,s}(k\text{ factor knot})/\mathbb{P}^\mathrm{H}_{\tube,s}(0_1)$, for $k=1,2,3$ in the $3\times1$ tube, together with straight line fits to the last few points (spans $\geq 1000$). (b) The corresponding plots for all polygons in the fixed-span model.}
\label{fig:3x1_1and2and3components_loglogfit}
\end{figure}

Figure~\ref{fig:3x1_1and2and3components_loglogfit} shows log-log plots of the probabilities of 1-, 2- and 3-factor knots, again divided by the probability of the unknot. For each we also include a straight-line fit to the last few points (spans $\geq 1000$). Note that for $k=1,2,3$, the slope of the line for the $k$-factor knots is very close to $k$ -- a further confirmation of the correctness of~\eqref{probKasymptotics}.

\subsection{Further evidence related to entropic exponents}

With the supposition that (based on the evidence from the previous section)
\begin{equation}
\mathbb{P}^{(\mathrm{sp})}_{\tube,s}(K) = A^{(\mathrm{sp})}_{\tube,K} s^{f_K}\left(\frac{\nu_{\tube,0_1}}{\nu_\tube}\right)^s(1+o(1))
\end{equation}
for a constant $A^{(\mathrm{sp})}_{\tube,K}$, we note that this quantity has a maximum at approximately
\begin{equation}
s \approx M_\tube(K) \equiv -\frac{f_K}{\log\left(\frac{\nu_{\tube,0_1}}{\nu_\tube}\right)}.
\end{equation}
(The exact location can depend on the $o(1)$ term since, for example, there are knot-types $K$ with minimum span in $\tube$ greater than $M_\tube(K)$.) An analogous form holds for Hamiltonian polygons. Note that $M_\tube(K)$ does not depend on the exact knot-type of $K$, only the number of prime knots in its decomposition. For the four models we focus on here, an approximate value for this location is given in Table~\ref{table:MK}.

\begin{table}[h]
\caption{Approximate values for $M_\tube(K)$ and $M^\mathrm{H}_\tube(K)$.}
\centering
\begin{tabular}{cll} \hline
	Tube & & \\
	Size &  $M_{\tube}(K)$ & $M_{\tube}^\mathrm{H}(K)$ \\
	\hline\hline
	$2\times1$		& $(1.23\times10^6)f_K$ & $34900f_K$	 \\
	\hline
	$3\times1$		& $7140f_K$ & $1400f_K$ \\
	\hline
\end{tabular}
\label{table:MK}
\end{table}

\begin{figure}
\centering
\begin{subfigure}{0.49\textwidth}
\includegraphics[width=\textwidth]{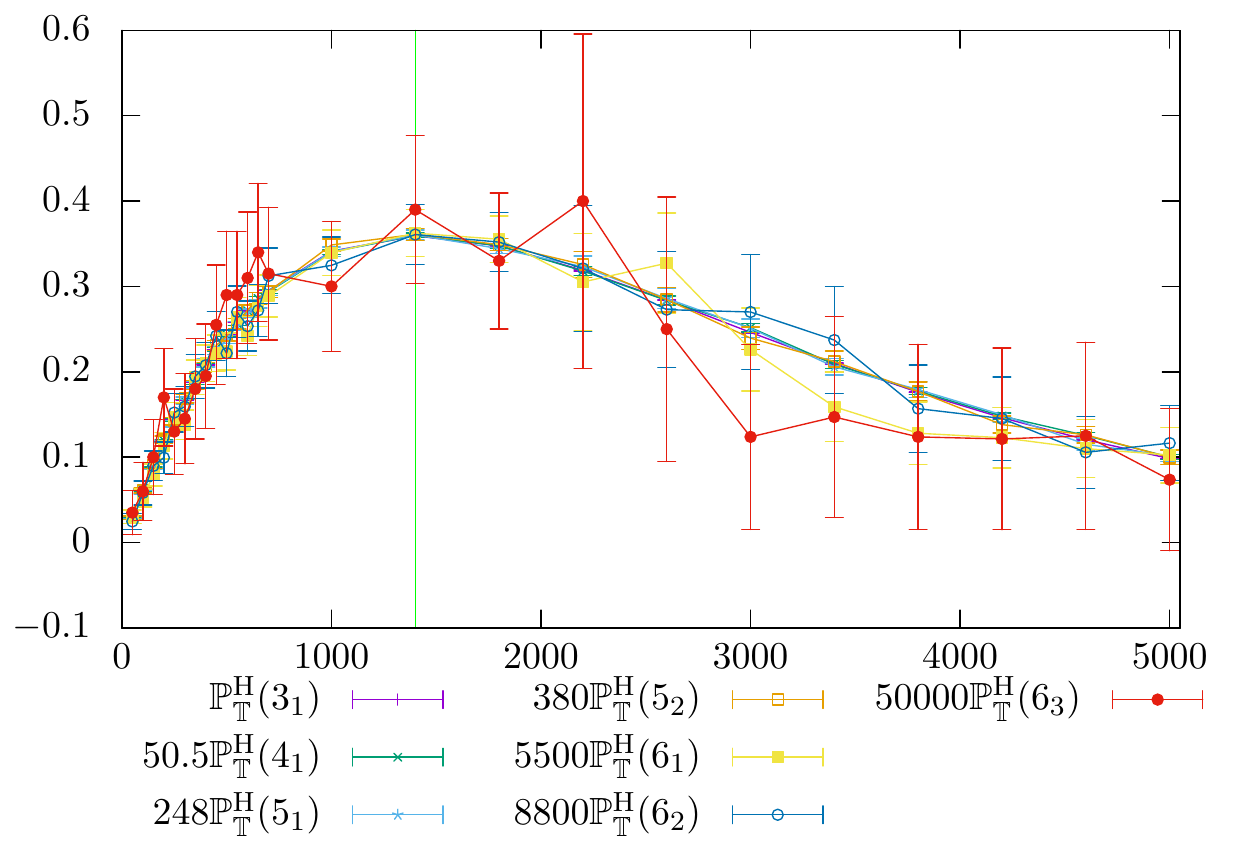}
\caption{}
\end{subfigure}
\begin{subfigure}{0.49\textwidth}
\includegraphics[width=\textwidth]{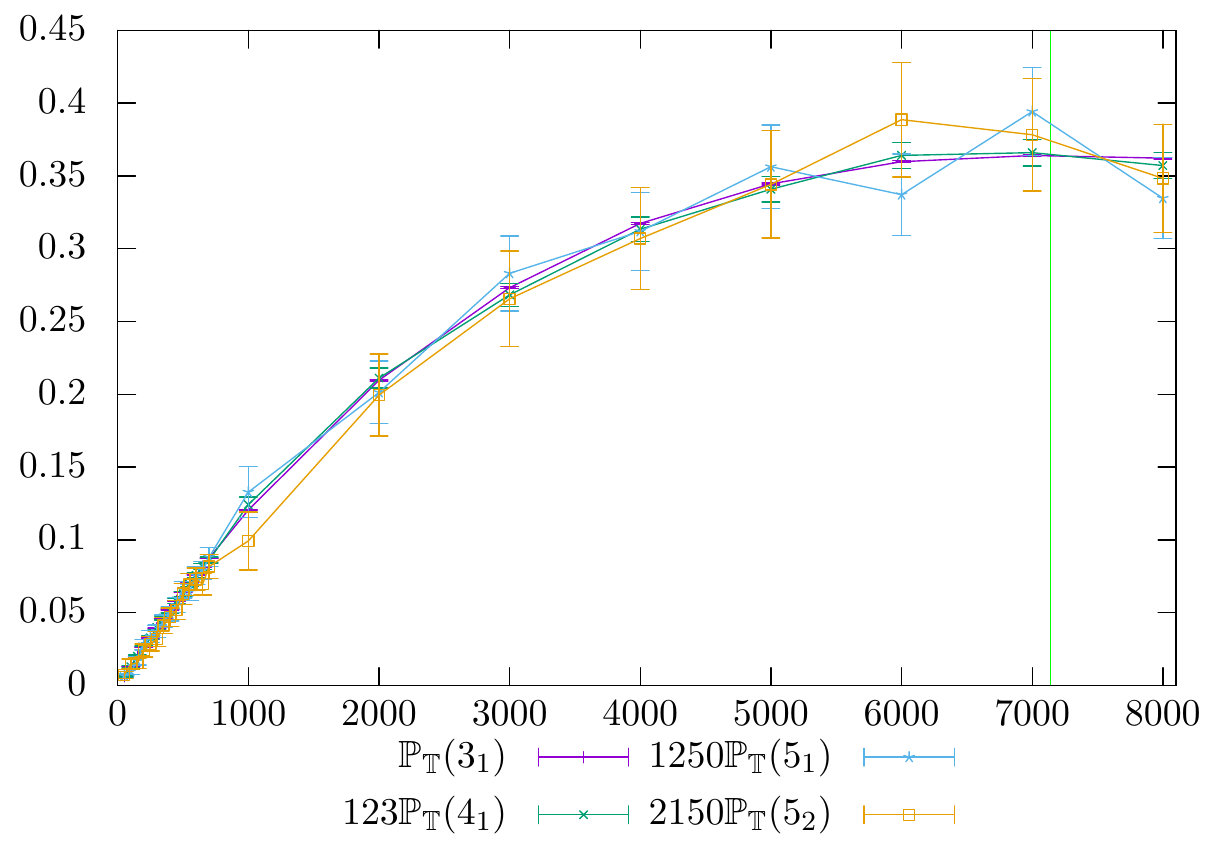}
\caption{}
\end{subfigure}
\caption{Plots of (a) $\mathbb{P}^\textrm{H}_{\tube}(K)$ and (b) $\mathbb{P}_{\tube}(K)$ for various prime knot types in the $3\times1$ tube, scaled by constant factors. The vertical lines at (a) 1400 and (b) 7140 indicate the predicted locations of the maxima, as per Table~\ref{table:MK}.}
\label{fig:3x1-prime_scaled}
\end{figure}

\begin{figure}
\centering
\begin{subfigure}{0.49\textwidth}
\includegraphics[width=\textwidth]{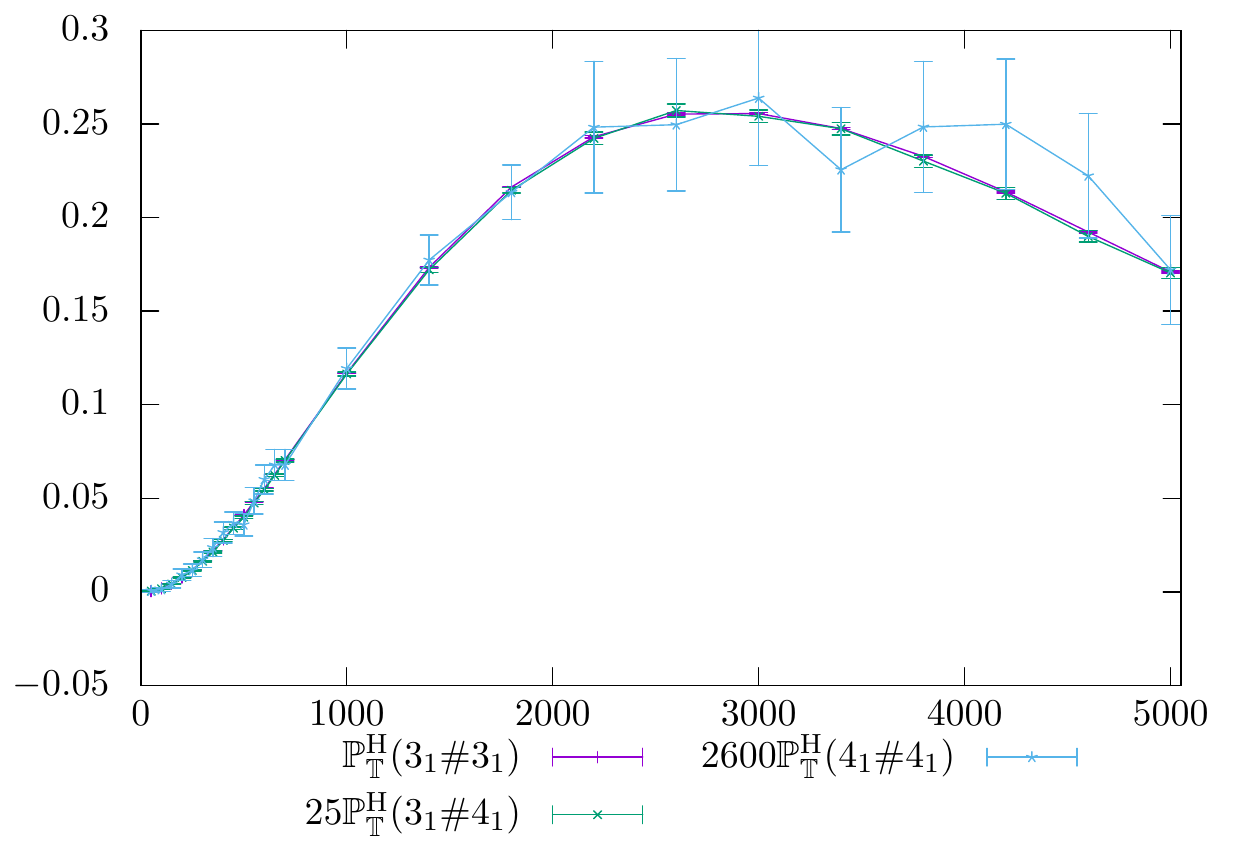}
\caption{}
\end{subfigure}
\begin{subfigure}{0.49\textwidth}
\includegraphics[width=\textwidth]{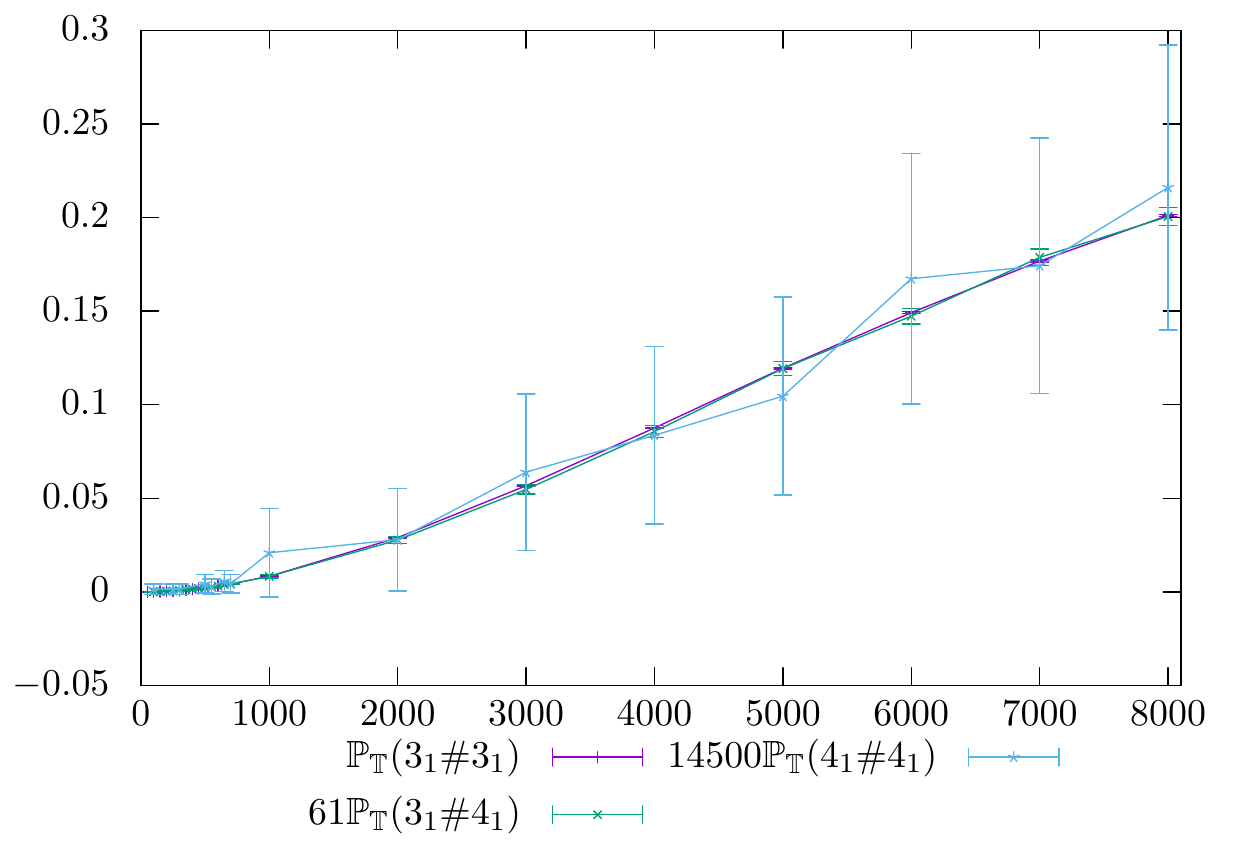}
\caption{}
\end{subfigure}
\caption{Plots of (a) $\mathbb{P}^\textrm{H}_{\tube}(K)$ and (b) $\mathbb{P}_{\tube}(K)$ for various 2-factor knot types in the $3\times1$ tube, scaled by constants.}
\label{fig:3x1-2comp_scaled}
\end{figure}

\begin{figure}
\centering
\begin{subfigure}{0.49\textwidth}
\includegraphics[width=\textwidth]{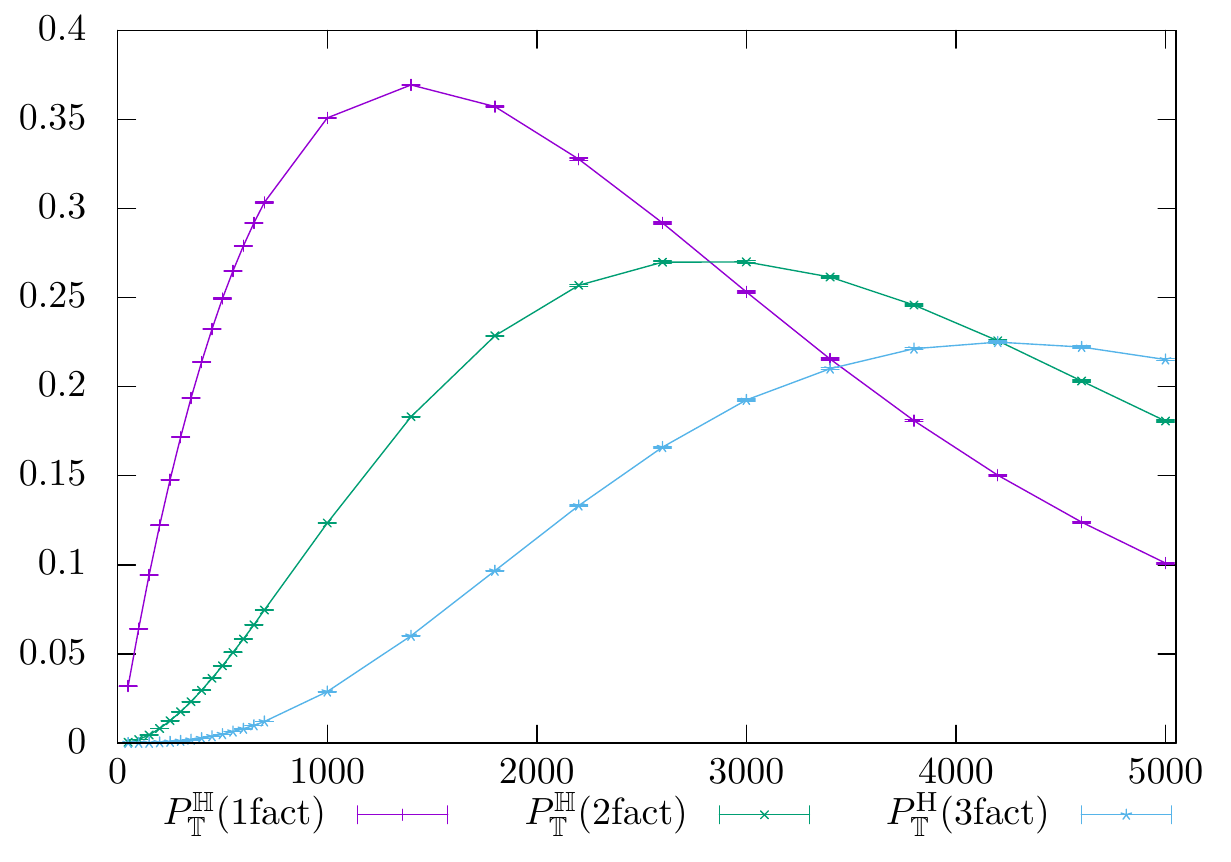}
\caption{}
\end{subfigure}
\begin{subfigure}{0.49\textwidth}
\includegraphics[width=\textwidth]{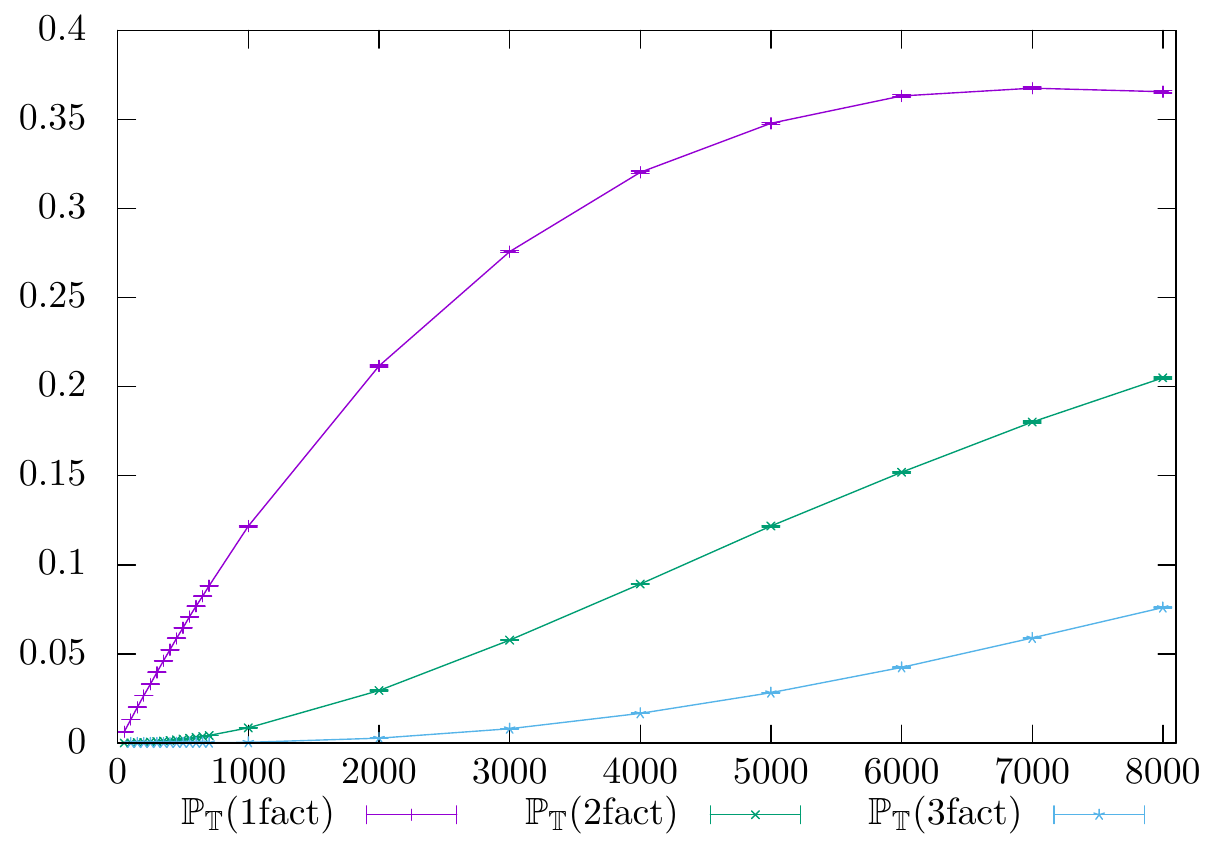}
\caption{}
\end{subfigure}
\caption{Plots of (a) $\mathbb{P}^\textrm{H}_{\tube}(k\text{ factor knot})$ and (b) $\mathbb{P}_{\tube}(k\text{ factor knot})$ for $k=1,2,3$ in the $3\times1$ tube.}
\label{fig:3x1-1and2and3comp_scaled}
\end{figure}

We have not attempted to sample polygons of sufficient span in the $2\times1$ tube to test the validity of the values given in Table~\ref{table:MK}. For the $3\times1$ tube, however, sufficiently long polygons have been generated to observe the maximum for $K=1$ (all polygons) and $K=1,2,3$ (Hamiltonian polygons). See Figures~\ref{fig:3x1-prime_scaled}--\ref{fig:3x1-1and2and3comp_scaled}. 

\subsection{Amplitude ratios}

We now briefly turn our attention to amplitude ratios for prime knot types. As mentioned in Section~\ref{sec:intro},  it is believed that the ratio $C_K/C_{3_1}$ (recall~\eqref{pnKasymptotics}) is universal, and does not depend on the lattice in question.

This cannot however be true when restricted to finite-size tubes, since for any given $L,M$ there will be infinitely many prime knot types $K$ which cannot be embedded in the $L\times M$ tube, so $C_K = 0$. (Indeed in \cite{Ishihara_2016} it has been established that only knots whose trunk is less than $(L+1)(M+1)$ are embeddable in an $L\times M$ tube.) This does however also suggest that even in tubes where knots of type $K$ can be embedded, we should not expect amplitude ratios to be universal, either between different tube sizes or between tubes and the full lattice.

To confirm this, in Table~\ref{table:ratios} we give some estimated amplitude ratios for a few different knot types in the $2\times1$ and $3\times1$ tubes, for both all and Hamiltonian polygons, as well as estimates from~\cite{1751-8121-44-16-162002} for the full lattice.

\begin{table}[h]
\caption{Estimates of amplitude ratios for 4- and 5-crossing prime knots in tubes. Insufficient data was available for useful estimates for all polygons in the $2\times1$ tube. Also included are estimates for the full lattice from~\cite{1751-8121-44-16-162002}.}
\centering
\begin{tabular}{llll} \hline
	Model &  $C_{4_1}/C_{3_1}$ & $C_{5_1}/C_{3_1}$ & $C_{5_2}/C_{3_1}$ \\
	\hline\hline
	$2\times1$ Ham.		& 0.0035  & 0.00011 & $8\times10^{-5}$ \\
	\hline
	$3\times1$ Ham.		& 0.020 & 0.0040 & 0.0026\\
	\hline
	$3\times1$ & 0.008 & 0.0008 & 0.00045 \\
	\hline
	Full lattice & 0.036 & 0.0025 & 0.0036 \\
	\hline
\end{tabular}
\label{table:ratios}
\end{table}

It seems reasonable to expect that as tube size increases, the amplitude ratios for knots in the tube, sampled from the fixed-edge ensemble, should approach the full lattice values. However, it is unclear what we should expect in the limit for the fixed-span or Hamiltonian ensembles.

\section{Conclusion}\label{sec:conclusion}

In this paper we have studied a model of self-avoiding polygons confined to a $L\times M$ tube of the cubic lattice $\mathbb Z^3$, and in particular considered the knotting properties of these polygons. This model is more tractable than that of polygons in the full lattice $\mathbb Z^3$, because it is characterised by a finite transfer matrix. Such a matrix characterisation allows, for example, to compute the generating function, growth rate and critical exponent for polygons in the tube. We have primarily focused on two particular ensembles of polygons: those enumerated by their span in the direction of the axis of the tube, and Hamiltonian polygons, which visit every vertex in an $L\times M \times s$ prism. This is primarily because knots are more common in these ensembles than for polygons enumerated by length.

However, polygons of fixed knot type (for example, unknots) cannot be characterised by a finite transfer matrix. For this reason we have developed a Monte Carlo algorithm for sampling random polygons directly from a chosen Boltzmann distribution. Using this algorithm we have been able to accurately estimate the growth rate for unknots and empirically confirm that the critical exponent for a given knot type $K$ corresponds to the number of prime factors in the knot decomposition of $K$. We have also estimated the sizes at which $k$-factor knots are most likely to occur, and verified these values using the Monte Carlo method. Finally, we have investigated amplitude ratios for given knot types in the tubes, and observed that these differ from those values (conjectured to be universal) in the full lattice.

The Monte Carlo method presented here has the distinct advantage over many other commonly used algorithms (e.g.~PERM, GAS, Wang-Landau, multiple Markov chain) in that it generates completely independent samples directly from the desired Boltzmann distribution. However, it also has a significant drawback -- the transfer matrix must be computed first, and must be held in memory during computation. Because the size of the transfer matrix grows extremely quickly with the size of the tube, this limits the method to only very small tube sizes.

Because polygons can be grown one ``slice'' at a time, a growth algorithm like PERM could be employed for this model.  We are currently investigating this and other sampling methods.

As mentioned in Section~\ref{sec:intro}, another work by the authors and collaborators~\cite{manyauthors_2x1tubes} is in preparation. This focuses exclusively on polygons in the $2\times1$ tube, and rigorously establishes~\eqref{pnKtubeinequality1}, i.e.~that the growth rates (counting by length) of certain knot types (2-bridge knots with unknotting number one, or connect sums thereof) are the same as that of the unknot, and moreover, the exponents of such polygon knots are indeed equal to the number of prime factors in the knot decompositions.

\section*{Acknowledgements}

NRB was supported by the PIMS Collaborative Research Group in Applied Combinatorics, and the Australian Research Council grant DE170100186. CES acknowledges support in the form of a Discovery Grant from NSERC (Canada) and a CPU allocation from Compute Canada's WestGrid. The authors also acknowledge assistance from Rob Scharein with KnotPlot and that some figures were produced using Rob Scharein's KnotPlot~\cite{knotplot}.

\sloppy
\printbibliography

\end{document}